\documentclass{aa}  
\usepackage{graphicx}
\usepackage{txfonts}
\usepackage{color}
\usepackage{natbib,twoopt}
\usepackage[hyphenbreaks]{breakurl}
\usepackage[breaklinks]{hyperref}                    
\bibpunct{(}{)}{;}{a}{}{,}            
\definecolor{cobalt}{rgb}{0.06, 0.2, 0.65}
\hypersetup{
  colorlinks,
  citecolor=cobalt,
  linkcolor=[rgb]{0.8, 0.2, 1.0},
  urlcolor=cobalt,
}
\makeatletter
  \newcommandtwoopt{\citeads}[3][][]{\href{http://adsabs.harvard.edu/abs/#3}%
    {\def\hyper@linkstart##1##2{}%
     \let\hyper@linkend\@empty\citealp[#1][#2]{#3}}}
  \newcommandtwoopt{\citepads}[3][][]{\href{http://adsabs.harvard.edu/abs/#3}%
    {\def\hyper@linkstart##1##2{}%
     \let\hyper@linkend\@empty\citep[#1][#2]{#3}}}
  \newcommandtwoopt{\citetads}[3][][]{\href{http://adsabs.harvard.edu/abs/#3}%
    {\def\hyper@linkstart##1##2{}%
     \let\hyper@linkend\@empty\citet[#1][#2]{#3}}}
  \newcommandtwoopt{\citeyearads}[3][][]%
    {\href{http://adsabs.harvard.edu/abs/#3}
    {\def\hyper@linkstart##1##2{}%
     \let\hyper@linkend\@empty\citeyear[#1][#2]{#3}}}
\makeatother

\renewcommand{\raggedright}{\leftskip=0pt \rightskip=0pt plus 0cm}

\begin{document} 

   \title{Molecular gas content of gravitational-lensed quasars at cosmic noon}

   \author{Zhiyuan Zheng \inst{1, 2, 5}
          \and
          Yong Shi \inst{1, 2, 3}\fnmsep\thanks{E-mail: yshipku@gmail.com}
          \and
          Qiusheng Gu \inst{1, 2}
          \and
          Zhi-Yu Zhang \inst{1, 2}
          \and
          Junzhi Wang \inst{4}
          \and
          Yanmei Chen \inst{1, 2}
          \and
          Fuyan Bian \inst{5}
          }

   \institute{
    School of Astronomy and Space Science, Nanjing University, Nanjing 210093, People's Republic of China\\
    \and
    Key Laboratory of Modern Astronomy and Astrophysics (Nanjing University), Ministry of Education, Nanjing 210093, People's Republic of China\\
    \and
    Department of Astronomy, Westlake University, Hangzhou 310030, Zhejiang Province, China\\
    \and
    Guangxi Key Laboratory for Relativistic Astrophysics, School of Physical Science and Technology, Guangxi University, Nanning 530004, PR China\\
    \and
    European Southern Observatory, Alonso de Córdova 3107, Casilla 19001, Vitacura, Santiago 19,Chile\\
    }


  \abstract{
   Star-forming activity in the host galaxies of high-redshift quasars is crucial to understanding the connection between supermassive black hole (SMBH) activity and galaxy evolution. While most existing studies are biased toward luminous quasars, we conduct carbon monoxide (CO) observations of 17 gravitationally lensed quasars that have four images using the IRAM 30m telescope to investigate the molecular gas content of moderate- to low-luminosity quasars. CO emissions are detected in five out of 17 quasars, corresponding to a detection rate of about 30\%. Analysis of their star formation activity reveals that these quasars live in gas-rich environments but exhibit weaker starbursts and lower star formation efficiencies compared to other luminous high-redshift quasars. In addition, the CO spectral line energy distributions of the two quasars (SDSS J0924+0219, SDSS J1330+1810) are also consistent with mild star formation instead of extreme starbursts. These results suggest that these lensed quasars reside in weaker starburst environments.} 
  

   \maketitle

\section{Introduction}

The modern framework for understanding the formation and evolution of galaxies emphasizes the importance of central supermassive black holes (SMBHs), which reveals a tight connection between SMBHs and their host galaxies \citep{Marconi03, Ho2008, Kormendy&Ho2013, Harrison2017}. The powerful winds or jets from accreting SMBHs, commonly called active galactic nuclei (AGN) feedback, play a significant role in the evolution of host galaxies, as stated by theoretical simulations \citep{Somerville2008, Crain2015}. Further, the global star formation history and black hole accretion history share a similar trend, peaking at redshift around two \citep{Shankar09, Madau14}, which is called the cosmic noon, also indicating the co-evolution scenario. The cosmic noon is the golden period for investigating the host galaxy properties of quasars, which facilitates the understanding of the interplay between AGNs and their host galaxies.

Quasars are the most luminous AGNs, and current knowledge of how they are triggered and grow is mainly through two mechanisms. Far-infrared (FIR) studies of quasars have revealed that luminous quasars tend to live in star-forming galaxies \citep{Shi2007, Shi14, Gurkan15, Stanley15, Harris16, Zhang16}  or gas- and dust-rich starbursting galaxies \citep{Omont01, Cox02, Pitchford16, Stacey2018, Salvestrini2025}. Theoretically, these quasars are thought to be triggered by gas-rich interactions or major mergers. A large amount of gas material feeds not only the black hole accretion \citep{Bergmann19} but also the extensive star formation \citep{Kennicutt1998, Bigiel2008}. This starburst-quasar framework is consistent with the living environment of luminous high redshift quasars \citep{Sanders1988, Alexander2005, Hopkins2008}. On the other hand, the discovery of a missing merger-AGN connection \citep{Grogin2005, Sharma2024} suggests that major mergers may not be the primary driver of AGN fueling. The continuous gas accretion might be another mechanism to ignite quasars \citep{Maccagni2014, Sabater2015, Tung2025}. A systematic analysis of merger-AGN connection across various AGN samples with different selection methods by \citet{Villforth2023} pointed out that the conflict between the weak or absence of merger-AGN connection and the starburst-quasar framework may be contributed not only by the sample bias but also the distinct physical conditions of the host galaxies, especially in low luminosity quasars. 

Therefore, it is a key to understanding the nature of host galaxies of intrinsic moderate- to low-luminosity quasars ($L_{\rm bol} \sim 10^{44} - 10^{46}\ {\rm ergs\;s^{-1}}$). Molecular gas clouds are sites where stars form. It is significant to study the star-forming activity by investigating the properties of the molecular gas content of quasar hosts. However, detections of molecular gas, mainly through carbon monoxide (CO) emission, are limited to high-luminosity quasar hosts with high star formation rate (SFR) and gas contents that are more similar to starburst galaxies \citep{Kakkad2017}. With boosted fluxes and spatial resolutions as offered by gravitational lensing, the lensed quasar provides a unique way to constrain the physical properties of host galaxies of moderate or low luminosity quasars at high redshift \citep{Blackburne2011}. Among all lensed quasars, those with four or more lensed images are treasures for the above purpose due to their fruitful information \citep{Sluse2003}. Until now, a total of 56 objects have been discovered (i.e., Quads\footnote{\url{https://research.ast.cam.ac.uk/lensedquasars/quads.html}}). The molecular gas observation is limited to part of these objects, and the results point to their hosts being more starburst-like galaxies (13/56, \citealp[]{Barvainis1997, Barvainis2002, Ao2008, Bradford2009, Riechers2011, Sluse2012, Deane2013, Paraficz2018, Stacey2020, Stacey2021, Stacey2022, Castillo2024}). To enlarge the sample size and probe star-forming main sequence (SFMS) -like quasar hosts, we conducted a molecular survey of the gravitationally-lensed quasars with four images from Quads using IRAM-30m telescope.

This paper is organized as follows. In \S~2, we introduce the IRAM 30m observations and the data reductions. In \S~3, we present the statistical features of our sample and the derived physical properties. In \S~4, we discuss the host galaxy properties of these quasars. Finally, we summarize our conclusions in \S~5. Throughout this work, the cosmological model is assumed as: $H_0 = 67.4$   $\rm km\;s^{-1}\;Mpc^{-1}$, $\Omega_{\rm m} = 0.315$ and $\Omega_{\rm \Lambda} = 0.685 $ \citep{PlanckCollaboration2020}.

\section{Observations and data reduction}

\begin{table*}
	\centering
        \caption{Observations conditions.}
        \label{table:obs}
	\begin{tabular}{l c c c c c c c} 
 
		\hline
            \hline
            Object & RA & DEC & ${\rm z_{qso}}$ & $\nu_{\rm obs}$(EMIR-band), transition & $\rm \Theta_{beam}$ & $\rm T_{sys}$ & $t_{\rm exp}$ \\
            & J2000 & J2000 &  & [GHz] & [\arcsec] & [K] & [hr] \\
            (1) & (2) & (3) & (4) & (5) & (6) & (7) & (8) \\ 
            \hline

            PSJ0147+4630    & 01:47:10.15 & 46:30:42.5  & 2.377 & 102.40(E0), J=3-2 & 24.1 &  75 & 11.0 \\
            SDSS J0924+0219 & 09:24:55.79 & 02:19:24.9  & 1.523 & 91.37(E0), J=2-1 & 26.8 & 107 & 8.4  \\ 
            SDSS J1330+1810 & 13:30:18.65 & 18:10:32.9  & 1.393 & 96.34(E0), J=2-1 & 25.5 &  91 & 18.8 \\ 
            H1413+117       & 14:15:46.24 & 11:29:43.4  & 2.56  & 97.13(E0), J=3-2 & 25.3 &  82 & 5.5  \\ 
            J2145+6345 & 21:45:05.11 & 63:45:41.2  & 1.56  & 90.05(E0), J=2-1 & 27.3 & 101 & 8.0  \\ 
            \hline

            PMNJ0134-0931 & 01:34:35.66 & -09:31:02.9 & 2.22  & 107.39(E0), J=3-2 & 22.9 &  85 & 7.4  \\ 
            COSMOS5921+0638 & 09:59:21.77 & 02:06:38.3  & 3.14  & 83.53(E0), J=3-2 & 29.4 & 119 & 8.7  \\ 
            SDSS J1004+4112 & 10:04:34.31 & 41:12:42.5  & 1.74  & 84.14(E0), J=2-1 & 29.2 &  87 & 8.0  \\
            J1042+1641 & 10:42:22.11 & 16:41:15.3  & 2.517$^\dagger$   & 98.80(E0), J=3-2 & 24.8 &  88 & 5.3 \\ 
            HE1113-0641     & 11:16:23.53 & -06:57:38.9 & 1.235 & 103.15(E0), J=2-1 & 23.8 & 134 & 7.4  \\ 
            SDSS J1138+0314 & 11:38:03.73 & 03:14:57.8  & 2.44  & 100.43(E0), J=3-2 & 24.4 & 105 & 11.5 \\ 
            HST14113+5211   & 14:11:19.61 & 52:11:29.7  & 2.811 & 90.74(E0), J=3-2 & 27.1 &  93 & 10.6 \\ 
            B1422+231       & 14:24:38.09 & 22:56:00.6  & 3.62  & 74.85(E0), J=3-2 & 32.8 &  92 & 11.5 \\
            SDSS J1433+6007 & 14:33:22.80 & 60:07:15.6  & 2.74  & 92.46(E0), J=3-2 & 26.6 & 102 & 6.0  \\
            J1817+2729 & 18:17:30.85 & 27:29:40.1  & 3.07  & 84.96(E0), J=3-2 & 28.9 &  95 & 7.6  \\ 
            B1933+503       & 19:34:30.90 & 50:25:23.2  & 2.64  & 95.05(E0), J=3-2 & 25.8 & 119 & 3.2  \\ 
            SDSS J2222+2745 & 22:22:09.50 & 27:45:33.8  & 2.82  & 90.52(E0), J=3-2 & 27.1 & 102 & 2.3  \\
            
		\hline
  
        \end{tabular} 
        \begin{flushleft}
            \textbf{Note:} (1) Object name. (2) Right Ascension. (3) Declination. (4) Redshift. (5) Observing CO transition. (6) Beam size (7) System temperature. (8) Total on-source exposure time.
            
            $\dagger$: the redshift of J1042+1642 was reported as 2.25 by \cite{Stacey2022}, while other observations, including X-ray, optical, and NIR spectra, all report the redshift of about 2.52 \citep{Matsuoka18, Walton2022, Glikman2023}. Therefore, we adopted 2.517 as the correct redshift measured from optical spectra of this object \citep{Glikman2023}.
        \end{flushleft}
        
\end{table*}

Our sample is a subset of quadruply image lensed quasars drawn from the Quads catalog, all classified as Type 1 quasars exhibiting broad emission lines with $\rm FWHM > 1200\ km\;s^{-1}$.
To maximize the observing efficiency, the declination of quasars is restricted to above -10$^{\circ}$. After excluding quasars with previous molecular gas observations until 2019, 24 quasars were selected for new molecular gas observations using the institut de radioastronomie millim\'{e}trique (IRAM) 30-meter telescope.

We carried out the molecular gas survey of these quasars in 2019 (project ID: 087-19, PI: Yong Shi), and 17 out of 24 objects have been observed. Table~\ref{table:obs} gives a summary of the basic information about our observations and targets. We observed CO J=2-1 and CO J=3-2 depending on the redshift of each object. The observations were carried out with the Eight Mixer Receiver (EMIR) in dual-polarization mode, using the Fourier Transform Spectrometers (FTS) backend, which provides a frequency resolution of 195 kHz and a coverage of 73 - 117 GHz in the rest frame. The standard wobbler switching (WSw) mode with a $\pm$120" offset at 0.5 Hz beam throwing was used for the observations \citep{Carter2012}. The average on-source integration time is 8.3 hours. Table~\ref{table:obs} summarizes the basic information of observations. 

Data reduction was conducted by the Continuum and Line Analysis Single-dish software (\texttt{CLASS} \footnote{\url{https://www.iram.fr/IRAMFR/GILDAS/}}). The stacked spectra were smoothed to a resolution of $\Delta V = 20.5\ {\rm km\;s^{-1}}$. Subsequently, we fitted the spectra over a window of $\pm 2500\ {\rm km\;s^{-1}}$ with a first-order polynomial for the baseline and a single Gaussian profile for the emission line, based on the redshift from the literature. This velocity range corresponds to the $\Delta z \sim \pm 0.019$ and $\pm 0.028$ for the CO J=2-1 and J=3-2 emission lines, respectively. To avoid the line missing due to the uncertainties of optical redshift, we also search for the emission lines across the full band, corresponding to  $\Delta z \sim \pm 0.29$ and $\pm 0.44$ for CO J=2-1 and CO J=3-2. However, since the typical uncertainty of optical redshift of quasars is about 0.02, corresponding to $< 1000\ \rm km/s$ \citep{Mazzucchelli2017}, the probability of missing the emission line due to imprecise optical redshift is negligible.

\section{Results}

\begin{table*}
        \tiny
        \caption{Estimated properties of the CO emissions.}
        \label{table:COprop}
	\begin{tabular}{l c c c c c c c c c } 
 
		\hline
            \hline
            Object & $ I_{\rm CO}$ & $ \Delta V_{\rm FWHM}$ & $ z_{\rm CO}$ & $ {\rm Log}\ \mu L_{\rm CO\ J1-0}^{'}$ & $\mu S_{\rm 160 \mu m}$ & $\rm Log\ \mu SFR$ & ${\rm Log}\ M_{\rm BH}$ & $\eta$ & $\mu_{\rm SF}$ \\ 
            
            & $\rm [K\;km\;s^{-1}]$ & $\rm [km\;s^{-1}]$ &  & $\rm [K\;km\;s^{-1}\;pc^2]$ & $\rm [mJy]$ & $[\rm M_{\odot}\;yr^{-1}]$ & ${\rm [M_{\odot}]}$ & &  \\ 
            (1) & (2) & (3) & (4) & (5) & (6) & (7) & (8) & (9) & (10) \\ 
            
            \hline

            PSJ0147+4630    & $ 0.75 \pm 0.07$ & $ 272 \pm 30$  & 2.3637 & $ 11.06 \pm 0.24$ & -- & -- & $10.05 \pm 0.35 ^{a}$ & $0.11 \pm 0.29 ^{a}$ & -- \\ 
            SDSS J0924+0219 & $ 0.65 \pm 0.12$ & $ 211 \pm 46$  & 1.5251 &  $11.0 \pm 0.28$ & $49.26 \pm 10.58$ & $3.31 \pm 0.28$ &  $7.93 \pm 0.34 ^{b}$ & $0.04 \pm 0.03 ^{b}$ & 23.5$^{c}$ \\ 
            SDSS J1330+1810 & $ 0.42 \pm 0.06$ & $ 219 \pm 36$  & 1.3945 & $ 10.75 \pm 0.26$ & $76.47 \pm 12.46$ & $3.40 \pm 0.27$ & $9.19 \pm 0.02 ^{d}$ & $0.02 \pm 0.02 ^{d}$ & $24 \pm 1$ $^{e}$ \\ 
            H1413+117       & $ 2.78 \pm 0.19$ & $ 528 \pm 42$  & 2.5585 & $ 11.68 \pm 0.23$ & $123.26 \pm 32.17$ & $4.26 \pm 0.28$ & $9.12 \pm 0.01  ^{b}$ & $0.28 \pm 0.06 ^{b}$ & 10.3$^{e}$ \\ 
            J2145+6345 & $ 0.54 \pm 0.09$ & $ 216 \pm 45$  & 1.5652 & $ 10.95 \pm 0.28$ & -- & -- & -- & -- & -- \\ 
            \hline
            
            PMNJ0134-0931 & $ < 0.92\ (3 \sigma) $ & 289 & -- & $ < 11.11\ (3 \sigma) $ & -- & -- & -- & -- & -- \\
            COSMOS5921+0638 & $ < 0.92\ (3 \sigma) $ & 289 & -- & $ < 11.35\ (3 \sigma) $ & -- & -- & -- & -- & -- \\ 
            SDSS J1004+4112 & $ < 0.87\ (3 \sigma) $ & 289 & -- & $ < 11.24\ (3 \sigma) $ & -- & -- & -- & -- & -- \\
            J1042+1641 & $ < 0.99\ (3 \sigma) $ & 289 & -- & $ < 11.22\ (3 \sigma) $ & -- & --  & $9.60 \pm 0.1 ^{f}$ & $0.05 \pm 0.01 ^{f}$ & $43 \pm 1 ^{e}$ \\  
            HE1113-0641     & $ < 1.68\ (3 \sigma) $ & 289 & -- & $ < 11.24\ (3 \sigma) $ & -- & -- & -- & -- & -- \\
            SDSS J1138+0314 & $ < 0.82\ (3 \sigma) $ & 289 & -- & $ < 11.12\ (3 \sigma) $ & $42.30 \pm 4.25$ & $3.75 \pm 0.26$ & $7.69 \pm 0.33  ^{b}$ & $0.43 \pm 0.36 ^{b}$ & -- \\
            HST14113+5211   & $ < 0.70\ (3 \sigma) $ & 289 & -- & $ < 11.16\ (3 \sigma) $ & $<36.38$ & $< 3.82$ & -- & -- & -- \\ 
            B1422+231       & $ < 0.66\ (3 \sigma) $ & 289 & -- & $ < 11.31\ (3 \sigma) $ & $7.96 \pm 0.88$ & $3.43 \pm 0.26$ & $9.72 \pm 0.38 ^{g}$ & $0.45 \pm 0.31 ^{g}$ & 24.22$^{h}$ \\ 
            SDSS J1433+6007 & $ < 1.22\ (3 \sigma) $ & 289 & -- & $ < 11.38\ (3 \sigma) $ & -- & -- & -- & -- & -- \\ 
            J1817+2729 & $ < 0.82\ (3 \sigma) $ & 289 & -- & $ < 11.29\ (3 \sigma) $ &  -- & -- & --  & -- & -- \\ 
            B1933+503       & $ < 1.57\ (3 \sigma) $ & 289 & -- & $ < 11.46\ (3 \sigma) $ & $160.95 \pm 22.3$ & $3.95 \pm 0.27$ & -- & -- & -- \\  
            SDSS J2222+2745 & $ < 1.63\ (3 \sigma) $ & 289 & -- & $ < 11.52\ (3 \sigma) $ & -- & -- & -- & -- & -- \\ 
            
		\hline  
        \end{tabular}
        
        \textbf{Note:} (1) Object name. (2) CO J=1-0 line luminosity converted from the high J-level emission line, assuming the line ratio as $R_{12}=1.2, R_{13}=1.8$ \citep{Tacconi2018}, which is not corrected by the gravitational lensing magnification. For non-detection objects, the flux is estimated to be three times the standard deviation of the continuum as the upper limit, and the full width at half maximum (FWHM) is assumed as the average of the five detected objects ($289 {\rm km\;s^{-1}}$). (3) FWHM of detected CO emission lines. (4) The inferred redshift from CO emission lines. (5)The inferred CO J1-0 luminosity without the magnification correction. (6) Flux density at 160 $\mu m$, which is inferred from the best fitted SED. (7) Star formation rate estimated from $160\ \mu m$ luminosity based on the eq.25 in \citet{Calzetti2010}. (8) The SMBH masses collected from the literature, which are corrected by the magnifications. (9) The Eddington ratio. (10) Magnification factors estimated from high-resolution CO emissions from the literature. Except B1422+231, whose magnification is estimated from mm continuum observations. 
        
        \textbf{Reference:}
        $^{a}$ This work, $^{b}$\citet{Sluse2012}, $^{c}$\citet{Badole2020}, $^{d}$\citet{Castillo2024}, $^{e}$\citet{Stacey2022}, $^{f}$\citet{Matsuoka18}, $^{g}$\citet{Assef2011}, $^{h}$\citet{Wen2022arXiv}
        
\end{table*}

\subsection{CO emission line detections}

\begin{figure*}
   \resizebox{\hsize}{!}
            {\includegraphics[width=\textwidth]{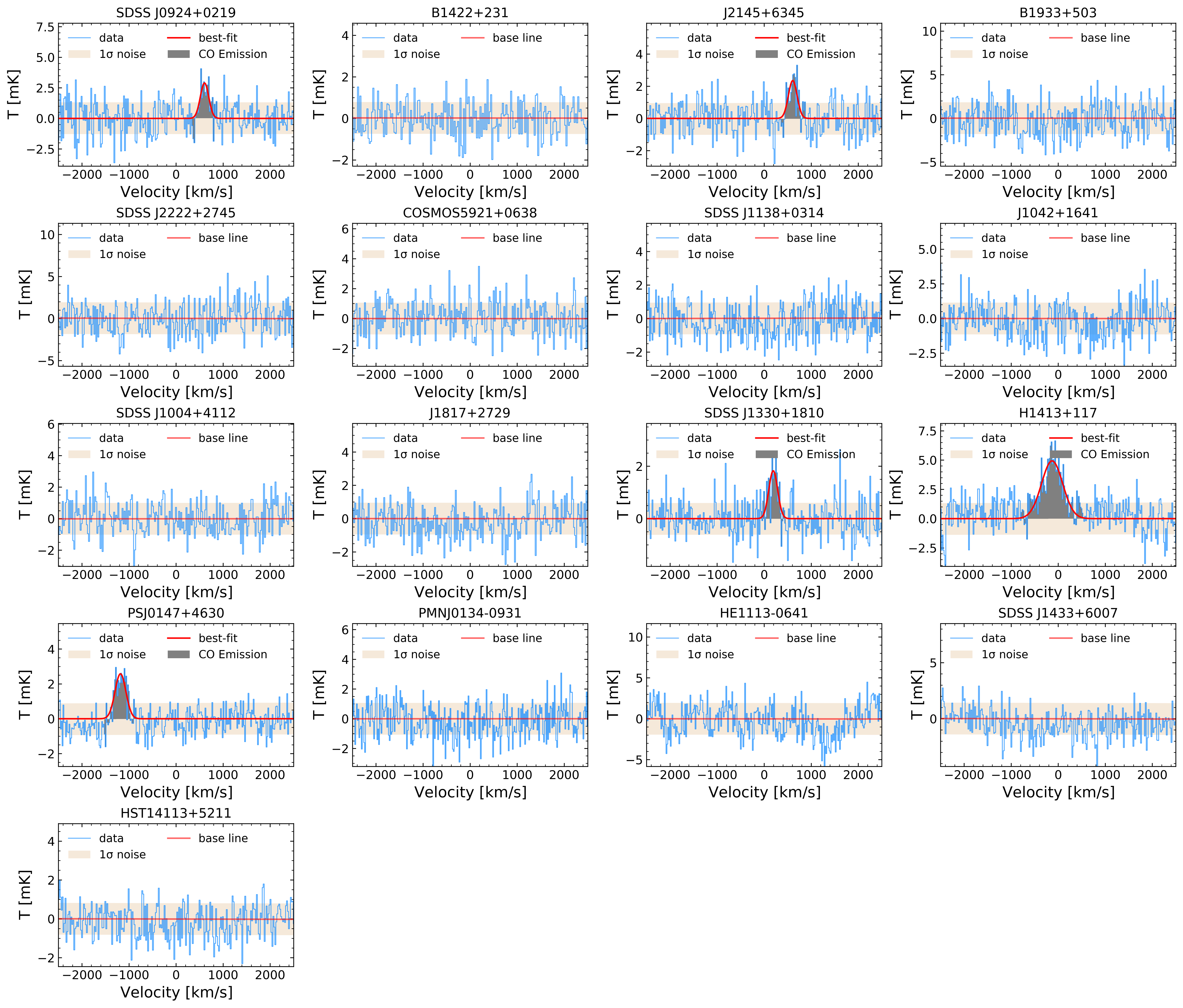}}
      \caption{Observed spectra at a velocity range from $\rm -2500\ km\;s^{-1}$ to $\rm 2500\ km\;s^{-1}$ of all objects, where the zero velocity is the optical redshift. The background blue curves are the stacked and smoothed spectra. The orange shadow represents the 1-$\sigma$ standard deviation of the continuum, while the grey region marks the CO emission line. The best-fitted Gaussian profile is marked by the red curve.}
         \label{fig:Spec}
\end{figure*}

Figure~\ref{fig:Spec} presents all the observed spectra of 17 objects. Five have reliable CO emission lines. The basic properties as listed in Table~\ref{table:COprop} are measured from the Gaussian profile. All these objects show significant line center velocity offset. We measure the redshift of CO emission lines ($z_{\rm CO}$) of quasars as compared to the redshift in the literature. The measured $z_{\rm CO}$ has significant offsets from the $z_{\rm opt}$ obtained from literature. For the three objects with previous CO observations, we compare our measured $z_{\rm CO}$ values with those reported in the literature and find them to be consistent.  Therefore, the offset is likely due to uncertainties in $z_{\rm opt}$ arising from broad line measurements. The CO luminosity is measured by the equation (2) from \citet{Solomon1997}. The upper limits to the CO luminosity of undetected objects are estimated as three times the continuum standard deviation by assuming a full width at half maximum (FWHM) of $\rm 289\ km\;s^{-1}$, the average of the five detected objects. 

The line ratios vary across different types of galaxies, quasars usually adopt a line ratio of $R_{\rm 12}=0.99$ and $R_{\rm 13}=0.97$ \citep{Carilli2013}. In contrast, the line ratios in star-forming galaxies are assumed as $R_{\rm 12}=1.2$ and $R_{\rm 13}=1.8$ \citep{Tacconi2018}. In this work, the final adopted line ratio is assumed to be the average of both, and the difference in line ratios is included in the uncertainties of the inferred CO J=1-0 line luminosity. In addition, a typical line flux-calibration uncertainty of about 10\% is also included in the total uncertainties. Other physical properties, such as SMBH masses and the Eddington ratio, are compiled from the literature and listed in the Table~\ref{table:COprop}. The magnification highly depends on the morphology of the emission traces. Since both the FIR and CO emissions are associated with the star-forming regions. The magnification of FIR and CO is expected to be consistent \citep{Ivison2002, Bussmann2013, Tuan2017, Stacey2018}. Therefore, for quasars without magnification estimates, the magnification factor ($\mu_{\rm SF}$) is assumed as $10^{+10}_{-5}$, consistent with the assumption for large samples of high redshift dusty star-forming galaxies \citep{Stacey2018}. For the quasars with previous CO high-resolution observations, the magnifications are adopted from the literature, as listed in Table~\ref {table:COprop}. The magnification of B1422+231 is estimated from ALMA 233 GHz continuum observations.

\subsection{CO spectral line energy distributions}

\begin{figure}[h]%
\centering
\includegraphics[width=9cm]{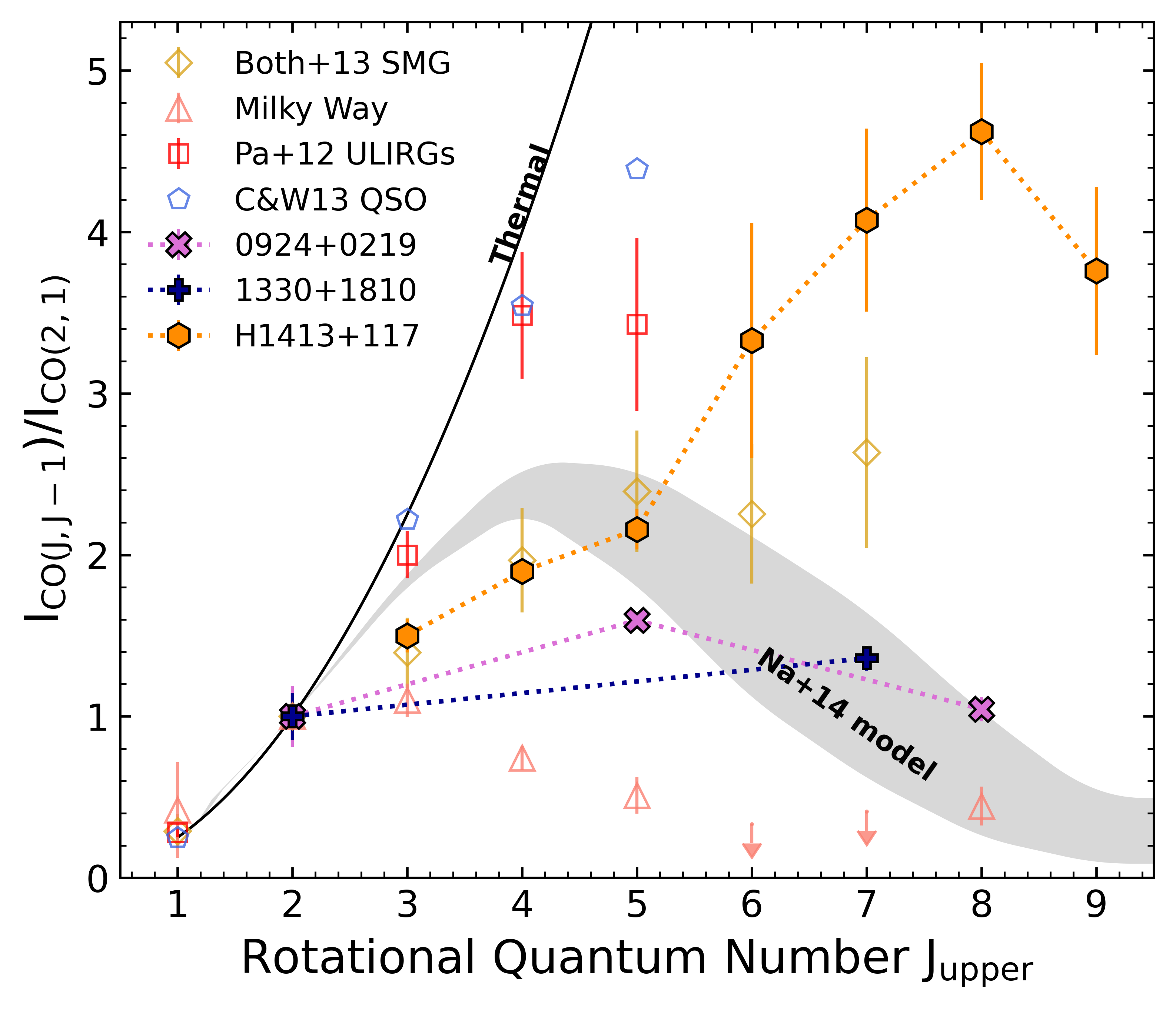}
\caption{CO spectral line energy distribution. The flux of other CO J ladders of our sample is collected from the literature. For J0924+0219, we collect the CO J=5-4 \citep{Badole2020} and J=8-7 \citep{Stacey2021}. For J1330+1810, we collect the CO J=7-6 \citep{Stacey2021}. For H1413+117, we collect the CO J=4-3, J=5-4, J=6-5, J=7-6, J=8-7, J=9-8 \citep{Barvainis1997, Bradford2009}. We compare our sample with several representative galaxy samples and theoretical simulations, including the inner disk of the Milky Way \citep{Fixsen1999}, the average SLED of local ULIRGs \citep{Papadopoulos2012}, SMGs \citep{Bothwell2013}, and luminous high-redshift quasars \citep{Carilli2013}, and the simulation-predicted SLEDs of star-forming galaxies with the $\Sigma_{\rm SFR} = 1-10 {\rm M_{\odot}\;yr^{-1}\;kpc^{-2}}$ \citep{Narayanan2014}. Except for H1413+117 (cloverleaf), other quasars show different SLED shapes, indicating distinct physical conditions of the interstellar medium within their host galaxies.}
\label{fig:SLED}
\end{figure}

The CO spectral line energy distributions (SLEDs) serve as powerful tracers of physical conditions of the molecular gas content, encoding information about gas densities and temperatures. The CO SLED of typical starburst-like galaxies at high-redshift peaks at higher J-level CO emission lines (about J=5) \citep{Weib2005, Panuzzo2010} than normal star-forming galaxies (about J=3-4) \citep{Fixsen1999, Daddi2015}. While for quasars, the CO is more highly excited due to their harder radiation field. Although the CO SLEDs of the two quasars are not complete, they still deviate from the typical starburst environments.

Combined with previously detected CO emissions from the literature, we draw the CO SLED for three quasars, SDSS J0924+0219, SDSS J1330+1810, and H1413+117 (Cloverleaf). For J0924+0219, we collect the CO J=5-4 \citep{Badole2020} and J=8-7 \citep{Stacey2021}. For J1330+1810, we collect the CO J=7-6 \citep{Stacey2022}. For H1413+117, we collect the CO J=4-3, J=5-4, J=6-5, J=7-6, J=8-7, J=9-8 \citep{Barvainis1997, Bradford2009}. The observed CO line temperature from IRAM 30m is converted to Jy by using the flux density to antenna temperature ratio S/T*A for a point source, with a value of 6 $\rm Jy/K$ for the 3-mm receiver. We compare the three quasars with several representative galaxy populations: the inner disk of the Milky Way \citep{Fixsen1999}, the average SLED of local ultra luminous infrared galaxies (ULIRGs) \citep{Papadopoulos2012}, SMGs \citep{Bothwell2013}, and luminous high-redshift quasars \citep{Carilli2013}. We also compared them with the simulation-predicted CO SLEDs of star-forming galaxies with the $\Sigma_{\rm SFR} = 1-10\ {\rm M_{\odot}\;yr^{-1}\;kpc^{-2}}$ \citep{Narayanan2014}. As shown in Figure~\ref{fig:SLED}, two of them show a low flux ratio at the high-J ladder, which deviates significantly from the luminous high-redshift quasars and the theoretical thermal limit, which suggests a sparse molecular environment and lower SFR density. The high-J CO flux is consistent with the theoretical prediction with a $\Sigma_{\rm SFR} = 1-10\ {\rm M_{\odot}\;yr^{-1}\;kpc^{-2}}$, which is far weaker than extreme starburst-like hosts ($\Sigma_{\rm SFR} \sim $ a few hundred ${\rm M_{\odot}\;yr^{-1}\;kpc^{-2}}$). While the CO SLED of Cloverleaf is consistent with the other high-redshift luminous quasar hosts, indicating a dense environment with high SFR density ($\Sigma_{\rm SFR} = 425\ {\rm M_{\odot}\;yr^{-1}\;kpc^{-2}}$, \citet{Solomon2003}). 

The magnification factor can vary across different CO emission lines, which primarily depends on the gas morphology traced by each line \citep{Sharon2019}. For J0924+0219 and J1330+1810, the high-J ladder CO emission was de-lensed by the magnification estimated from the corresponding spatially-resolved CO emission line maps. Although slight variations in magnification may occur between different CO lines \citep{Sharon2019}, such differences in the magnification are not expected to affect the shape of the normalized CO SLEDs significantly. We therefore assume a constant magnification across different CO transitions.

\subsection{Star formation rate estimation}

To avoid the significant AGN contamination in the total IR luminosity, which might lead to an overestimate of star formation rate, we estimated the star formation rate from the $160\ \mu m$ luminosity (eq.25 in \citealt{Calzetti2010}), a wavelength less affected by the AGN heating \citep{DiMascia2023}. The $160\ \mu m$ luminosity was obtained by performing empirical SED fitting with joint AGN SED templates \citep{Dale2014} and interpolating the best-fit SED to $160\ \mu m$ in log–log space. For quasars and ULIRGs in Figure~\ref{fig:SF}, we applied the above SED fitting and estimated their SFRs. The associated uncertainties include the systematic scatter of the calibration ($\sim 0.4\ \rm dex$) and those propagated from the $160\ \mu m$ luminosity via MCMC. Photometric data were compiled from literature, i.e., the lensed quasars in our sample and Quads samples from \citet{Stacey2018}, and reference therein, PG quasars from \citet{Shi2014, Shangguan18}, high-redshift quasars from \citet{Solomon2005, Riechers2006, Circosta2021}, ULIRGs from \citet{Solomon1997}. For other star-forming samples and the SFMS relation, SFRs were derived from IR luminosity ($8-1000\ \mu m$) using eq.4 of \citet{Kennicutt1998}. The inferred $160\ \mu m$ flux densities and SFRs for our samples are listed in Table~\ref{table:COprop}, without magnification corrections.

\section{Discussion}

\subsection{Star-forming activity of quasar hosts}

\begin{figure*}
   \resizebox{\hsize}{!}
            {\includegraphics[width=\textwidth]{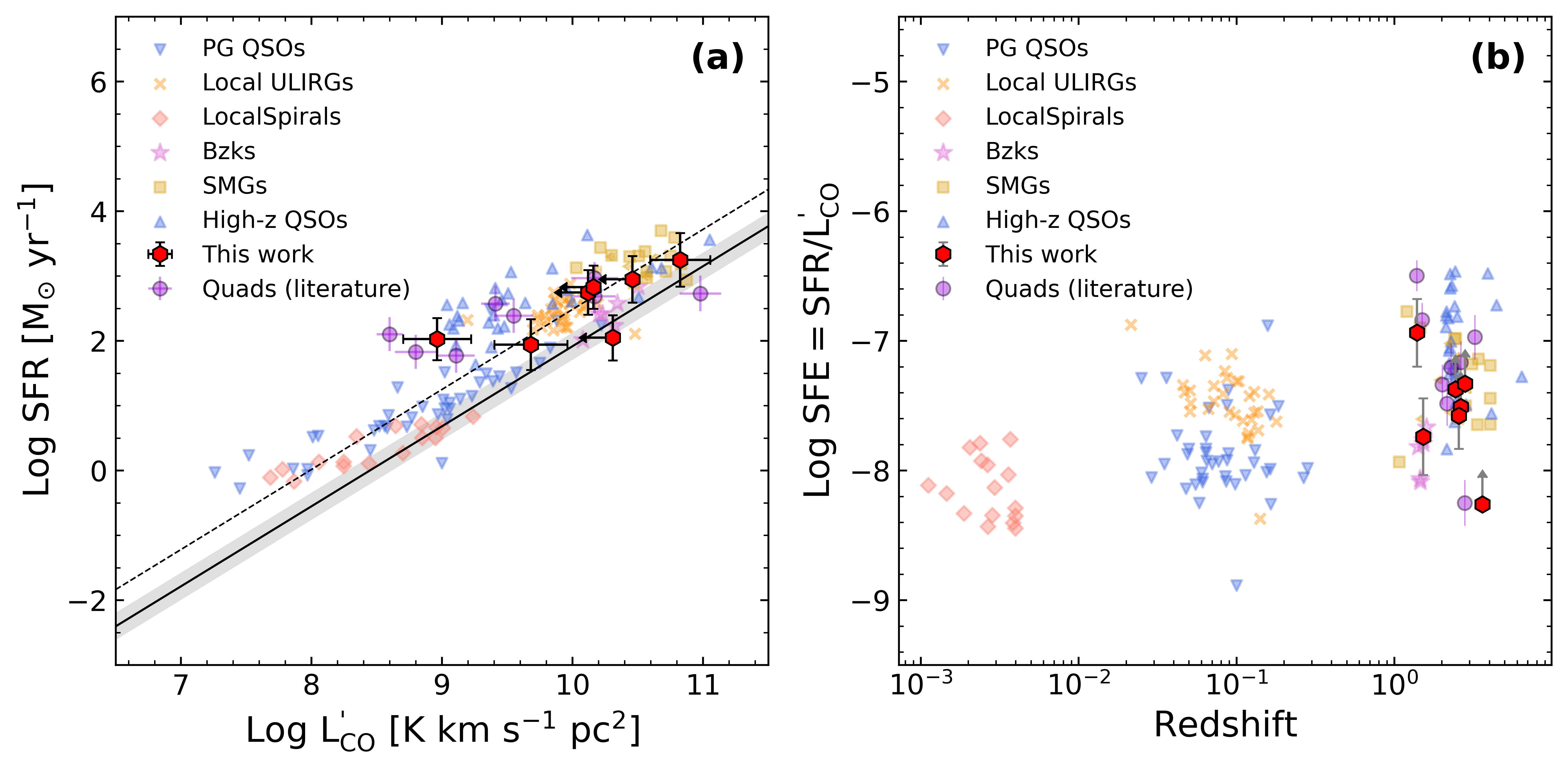}}
      \caption{Comparison of star-forming activity between our sample and other representative galaxy samples. (a) CO J=1-0 vs. SFR. The black solid line shows the star-forming main sequence (SFMS) with 1-$\sigma$ scatter, while the black dotted line shows the starburst trend \citep{Sargent2014}. (b) Star formation efficiency (${\rm SFE} \equiv {\rm SFR} / L_{\rm CO\ J=1-0}^{'}$) as a function of redshift. Various galaxy samples include the near-infrared selected (Bzk) galaxies \citep{Daddi2010}, SMGs \citep{Greve2005, Daddi2009a, Daddi2009b}, luminous high-redshift quasars \citep{Solomon2005, Riechers2006}, local PG quasars \citep{Shangguan18, Shangguan2020}, local ULIRGs \citep{Solomon1997}, local spirals \citep{Leroy2008, Leroy2009, Wilson2009}, and Quads lensed quasars from literature \citep{Barvainis1997, Barvainis2002, Ao2008, Bradford2009, Riechers2011, Deane2013, Paraficz2018, Stacey2020, Stacey2021, Stacey2022, Castillo2024}.}
         \label{fig:SF}
\end{figure*}

\begin{figure}[h]%
    \centering
    \includegraphics[width=9cm]{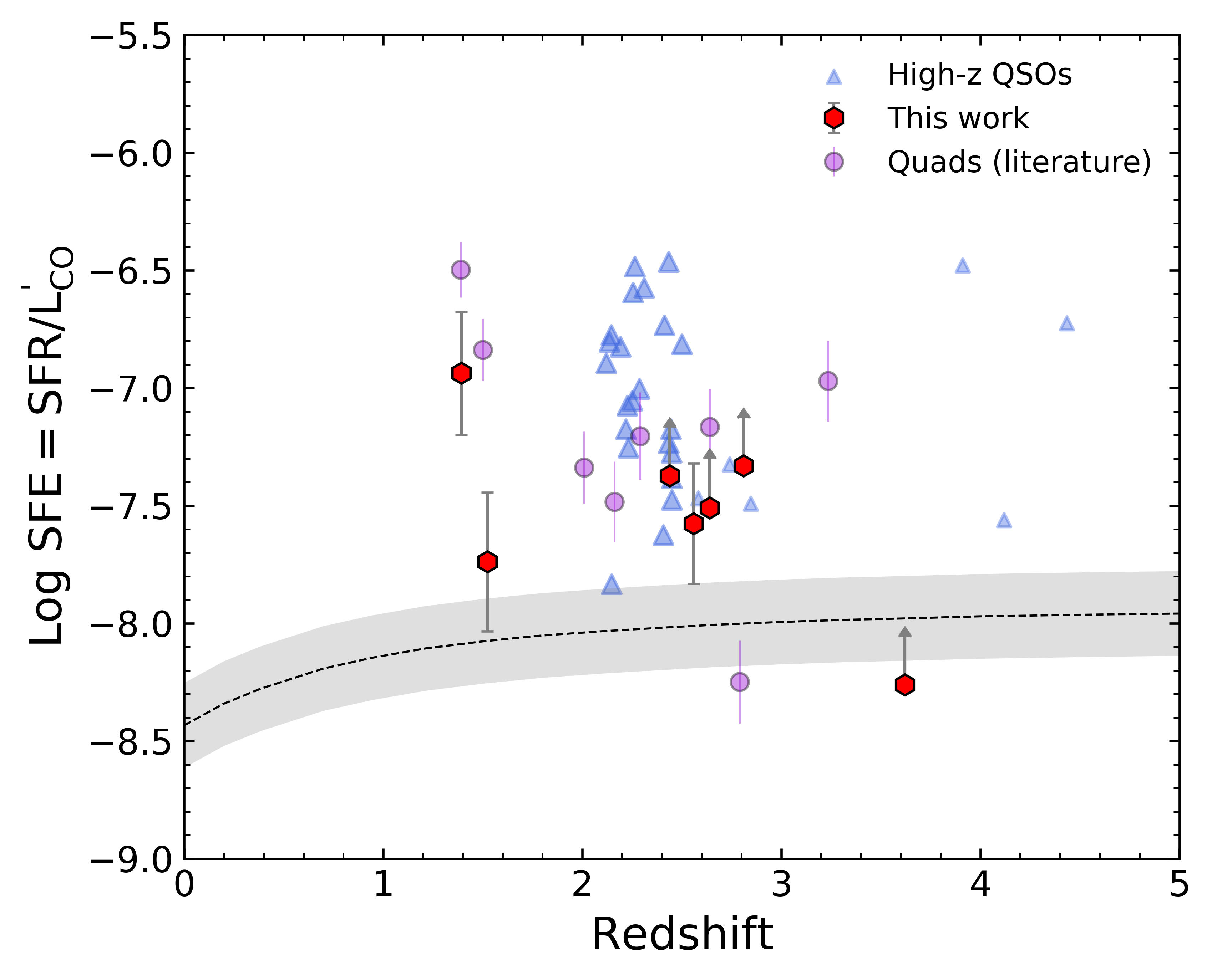}
    \caption{SFE as a function of redshift, adding the inferred SFE cosmic evolution of star-forming main sequence from \citet{Sargent2014} as indicated with the black dotted line with 1-$\sigma$ scatter. Compared to other high-z quasars \citep{Solomon2005, Circosta2021}, the host galaxies of our quasars show a lower SFE.}
    \label{fig:SFEzMS}
\end{figure}

The CO luminosity manifests strong correlations with total IR  luminosity ($8-1000\ \mu m$) across both the local and distant universe \citep{Sanders1985, Solomon2005, Carilli2013, Sargent2014}, which show different trends between starburst and SFMS galaxies. To minimize the effect of the different SED fitting methods and AGN contamination on the total IR luminosity, we replaced the IR luminosity with the FIR-based SFR. Figure~\ref{fig:SF} (a) displays the SFR as a function of CO J=1-0 luminosity. We compare our samples with the best-fitted SFMS and starburst relation galaxies at $0<z<3$ \citep{Sargent2014} and other representative galaxy samples, including the local spiral galaxies \citep{Leroy2008, Leroy2009, Wilson2009}, local ULIRGs \citep{Solomon1997}, local PG quasars (\citealp{Shangguan18, Shangguan2020}, and reference therein), high-redshift Type 1 quasars \citep{Solomon2005, Riechers2006, Circosta2021}, SMGs \citep{Greve2005, Daddi2009b, Daddi2009a}, near-infrared selected (Bzk) galaxies \citep{Daddi2010}, and Quads lensed quasars from literature \citep{Barvainis1997, Barvainis2002, Ao2008, Bradford2009, Riechers2011, Deane2013, Paraficz2018, Stacey2020, Stacey2021, Stacey2022, Castillo2024}. For star-forming galaxy samples, the SFRs are estimated using eq.4 of \citet{Kennicutt1998}, while for other quasar samples, the SFR is derived from the $160\ \mu m$ luminosity as described in \S~3.3. Our sample spans about two orders of magnitude in both SFR and CO luminosity. These quasars follow the starburst sequence but, on average, show weaker starburst activity than other high-redshift quasar samples.

The ratio between the SFR and CO luminosity serves as a proxy for star formation efficiency, defined as ${\rm SFE} \equiv {\rm SFR} / L_{\rm CO\ J=1-0}^{'}$ in unit of ${\rm 10^{-9}\ M_{\odot}\;yr^{-1}/(K\;km\;s^{-1}\;pc^2)}$. Figure~\ref{fig:SF} (b) compares the SFE of our objects with the same samples in Figure~\ref{fig:SF} (a) across cosmic time. These quasars exhibit relatively low SFEs, ranging from $5.5 - 46.6$ with a median of $\sim 30$ (including upper limits). This value is higher than that of normal star-forming galaxies (median $\sim 7$) and local PG quasars (median $\sim 12$), but lower than that of other high-redshift quasars (median  $\sim 65$). Although we employed a different SFR estimation method between the purely star-forming galaxies and quasars, the potential systematic bias ($\sim 0.3\ \rm dex$) does not significantly affect our conclusions. Meanwhile, the CO SLED of two quasars (SDSS J0924+0219 and SDSS J1330+1810) are also consistent with the simulation of less starburst environments. These results suggest that the host galaxies of these quasars experience less starburst activity than other high-redshift quasars. 

Since the unresolved CO detections alone are insufficient to estimate the magnified factor ($\mu_{\rm SF}$) of our quasars, we adopt the estimated $\mu_{\rm SF}$ from the previous spatially resolved CO observations in the literature. For quasars without estimation, we adopt a magnification factor of $\mu_{\rm SF} = 10_{-5}^{+10}$. It has been reported that the magnification can vary even a few times with wavelength/frequency \citep{Deane2013, Zhang2023}. However, it is mainly due to the distinct geometry of structures. In terms of the magnification estimated from FIR and CO observation, which both trace the star-forming regions, the magnification should remain consistent. On the other hand, the magnification affects the SFR and CO luminosity simultaneously, resulting in only minor effects on the SFEs and the offset from the SFMS ($\Delta$MS). Consequently, the uncertainties in gravitational lensing magnification only have a minor effect on the $\Delta$MS, which will not bias our conclusion.

\subsection{The SFEs versus quasar properties }

\begin{figure}[h]%
    \centering
    \includegraphics[width=9cm]{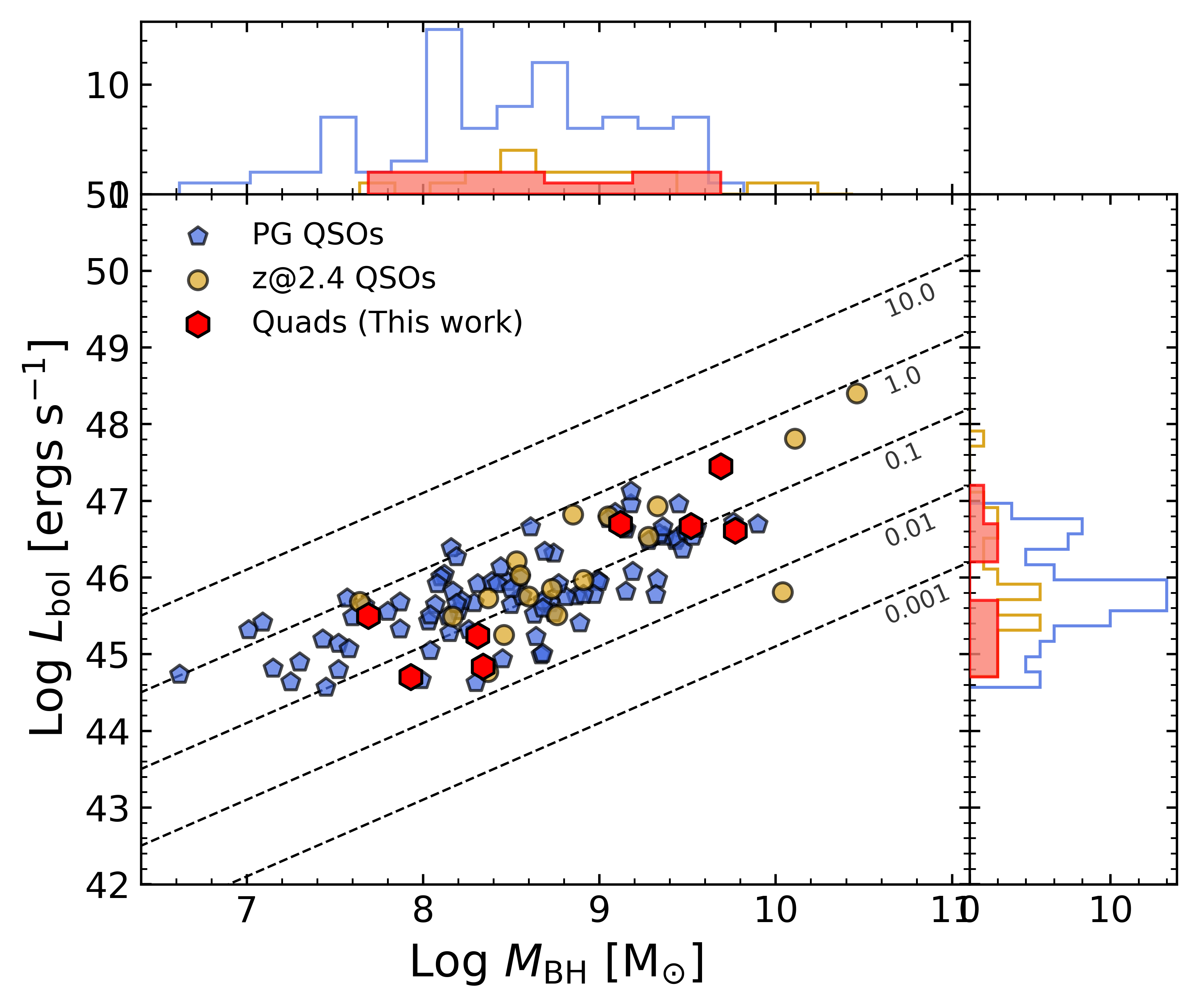}
    \caption{Comparison of the SMBHs mass and bolometric luminosity with local \citep{Shangguan18} and distant quasars \citep{ Circosta2021}. The dashed lines represent the constant Eddington ratios. The black hole masses and bolometric luminosities are corrected with magnification. The black hole mass of our sample is distributed between the local PG quasars and luminous quasars at cosmic noon, while bolometric luminosities show a similar distribution.}
    \label{fig:MBHEdd}
\end{figure}

\begin{figure}[h]%
    \centering
    \includegraphics[width=9cm]{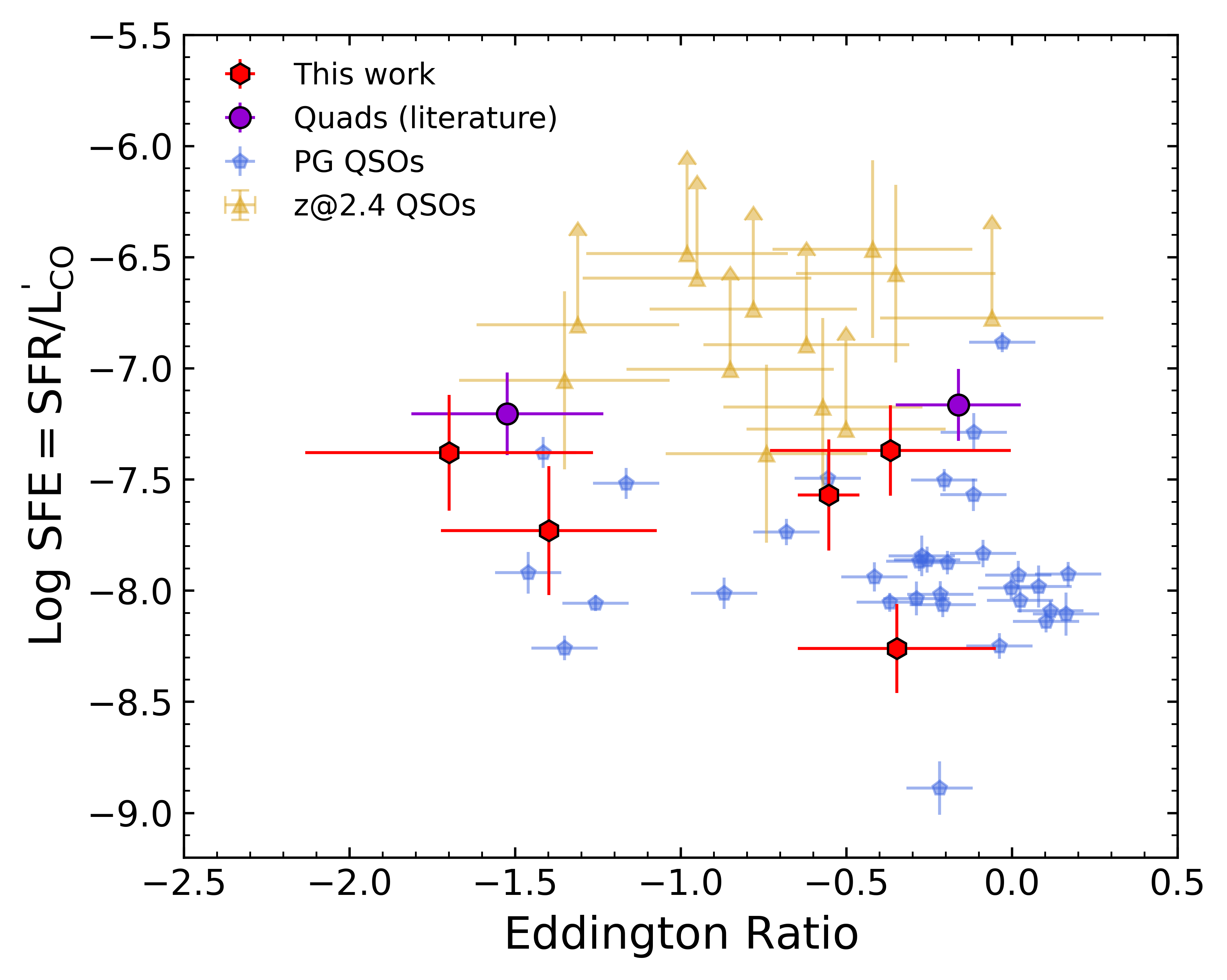}
    \caption{Comparison of the SFE and the Eddington ratio with local \citep{Shangguan18, Shangguan2020} and distant quasars \citep{Circosta2021}. The results show the SFE of our quasars is lower than that of high-redshift quasars but comparable to local PG quasars, while the Eddington ratio is slightly lower than that of other samples. }
    \label{fig:SFEEdd}
\end{figure}

Figure~\ref{fig:MBHEdd} illustrates the comparison of basic quasar properties between our sample with both local PG quasars \citep{Shangguan18, Shangguan2020} and luminous high-redshift quasars \citep{Circosta2021, Stacey2018}. Notably, PSJ0147+4630 is a broad absorption line quasar, making it difficult to estimate the SMBH mass through normal broad lines such as [C IV] and Mg II \citep{Lee2017}. Therefore, the SMBH mass is estimated from the C III] broad line in this work, which introduces large uncertainties in both the SMBH mass ($\sim 0.5$ dex) and the Eddington ratio (\citealp{Popovic2020}, and reference therein). For other quasars, the $M_{\rm BH}$ is adopted from literature as listed in Table~\ref{table:COprop}. While bolometric luminosities are estimated from monochromatic luminosity using the bolometric correction of \citet{Shen2008}. The BH mass of our quasars falls between those of local and high-redshift quasars, while the bolometric luminosity shows a similar distribution. About half of our objects (4/7) are moderate-luminosity (${\rm log}\ L_{\rm bol} < 46\ {\rm ergs\;s^{-1}}$), and both the lower SMBH mass and Eddington ratio are responsible for their moderate luminosity. In Figure~\ref{fig:SFEEdd}, we further compared the Eddington ratio and the SFE of their host galaxies with other quasar samples. Although our quasars exhibit sub-Eddington accretion, their host galaxies display less starburst, which is milder than that of other high-redshift quasars. The physical condition of our quasar hosts might differ from that of luminous quasars at high redshift. 

\subsection{The weaker starburst hosts of high-redshift quasars}

The properties of molecular gas content in high-redshift quasar host galaxies help us to reveal their living environment, and further facilitate our understanding of how black hole accretion shapes the evolution of galaxies. In the typical ignition scenario for high-redshift quasars, there exists an evolutionary phase during which quasar feedback becomes energetic enough to eliminate the dust and gas from the galactic center, and unveil the central SMBHs \citep{Kennicutt1998, Sanders1988, Hopkins2008, Maccagni2014, Lapi2018, Villforth2023, Tung2025}. This leads to a low accreting efficiency pattern of quasars, which is characterized by low accretion efficiency and mild star-forming activity within the host galaxy. This evolution stage is far supported by local quasars, ranging from Seyfert I and II galaxies \citep{Husemann2017, Salvestrini2022} to PG quasars \citep{Zhang16, Shangguan2020, Molina2023}. Local studies reveal the two distinct distributions in terms of $L_{\rm CO}^{'}/L_{\rm IR}$, large amounts of Seyferts reside at the SFMS with relatively lower luminosity \citep{Koss2021}, while more luminous PG quasars are aligned with the starburst trend consistent with high-redshift quasars \citep{Molina2023}. However, quasars living in less starburst hosts are difficult to observe at high redshift due to their intrinsic low CO luminosity. Benefiting from the gravitational lensing, the host galaxies of these quasars are living in less starburst systems, as illustrated in Figure~\ref{fig:SF} and ~\ref{fig:SFEzMS}.

\section{Conclusions}

We conducted a molecular gas survey of 17 gravitational lensed quasars with four images selected from Quads using the IRAM-30m telescope, leveraging gravitational lensing magnification to probe the CO emission of intrinsically moderate/low luminosity quasars at high redshift. As a result, the CO emissions are detected in five out of 17 quasars ($\sim 30\%$). By combining archival photometric data and SMBH properties from the literature, we compared the star formation activity of their host galaxies and the SMBH masses, Eddington ratio, with several typical galaxy samples in both the local and distant universe. The following are the main conclusions:

\begin{enumerate}

    \item The five quasars with CO detections reside in gas-rich environments, with a median CO J=1-0 luminosity of about $\rm 10^{10.1}\ K\;km\;s^{-1}\;pc^{2}$ after the correction of magnification of gravitational lensing. 

    \item For the quasars with sufficient archival photometric data to perform SED fitting, their host galaxies exhibit weaker starburst and relatively lower star formation efficiency, with a median value of ${\sim 30 \times \rm 10^{-9}\ M_{\odot}\;yr^{-1}/(K\;km\;s^{-1}\;pc^2)}$, roughly half that of the comparison sample of high-redshift quasars.

    \item After collecting the detection of other CO emission lines, we draw the CO SLED of three quasars: SDSS J0924+0219, SDSS J1330+1810, and H1413+117. The CO SLED of the first two quasars is more consistent with the simulation of less starburst galaxies, while H1413+117 shows an extremely luminous quasar-like CO SLED. 
    
   \end{enumerate}

\begin{acknowledgements}

We would like to thank the referee for their valuable suggestions and comments, which significantly helped improve this work. 
This work is based on observations carried out under project number 087-19 with the IRAM 30m telescope. IRAM is supported by INSU/CNRS (France), MPG (Germany) and IGN (Spain). This work acknowledges the support from the National Key R\&D Program of China No. 2023YFA1608204, the National Natural Science Foundation of China (NSFC grants 12141301, 12121003, 12333002).
        
\end{acknowledgements}

\bibliographystyle{aa_url.bst}
\bibliography{aa55046-25}

\begin{thebibliography}{97}
\expandafter\ifx\csname natexlab\endcsname\relax\def\natexlab#1{#1}\fi

\bibitem[{{Alexander} {et~al.}(2005){Alexander}, {Smail}, {Bauer}, {Chapman}, {Blain}, {Brandt}, \& {Ivison}}]{Alexander2005}
{Alexander}, D.~M., {Smail}, I., {Bauer}, F.~E., {et~al.} 2005, \href{http://dx.doi.org/10.1038/nature03473}{\color{magenta}\nat}, \href{https://ui.adsabs.harvard.edu/abs/2005Natur.434..738A}{434, 738}

\bibitem[{{Ao} {et~al.}(2008){Ao}, {Wei{\ss}}, {Downes}, {Walter}, {Henkel}, \& {Menten}}]{Ao2008}
{Ao}, Y., {Wei{\ss}}, A., {Downes}, D., {et~al.} 2008, \href{http://dx.doi.org/10.1051/0004-6361:200810482}{\color{magenta}\aap}, \href{https://ui.adsabs.harvard.edu/abs/2008A&A...491..747A}{491, 747}

\bibitem[{{Assef} {et~al.}(2011){Assef}, {Denney}, {Kochanek}, {Peterson}, {Koz{\l}owski}, {Ageorges}, {Barrows}, {Buschkamp}, {Dietrich}, {Falco}, {Feiz}, {Gemperlein}, {Germeroth}, {Grier}, {Hofmann}, {Juette}, {Khan}, {Kilic}, {Knierim}, {Laun}, {Lederer}, {Lehmitz}, {Lenzen}, {Mall}, {Madsen}, {Mandel}, {Martini}, {Mathur}, {Mogren}, {Mueller}, {Naranjo}, {Pasquali}, {Polsterer}, {Pogge}, {Quirrenbach}, {Seifert}, {Stern}, {Shappee}, {Storz}, {Van Saders}, {Weiser}, \& {Zhang}}]{Assef2011}
{Assef}, R.~J., {Denney}, K.~D., {Kochanek}, C.~S., {et~al.} 2011, \href{http://dx.doi.org/10.1088/0004-637X/742/2/93}{\color{magenta}\apj}, \href{https://ui.adsabs.harvard.edu/abs/2011ApJ...742...93A}{742, 93}

\bibitem[{{Badole} {et~al.}(2020){Badole}, {Jackson}, {Hartley}, {Sluse}, {Stacey}, \& {Vives-Arias}}]{Badole2020}
{Badole}, S., {Jackson}, N., {Hartley}, P., {et~al.} 2020, \href{http://dx.doi.org/10.1093/mnras/staa1488}{\color{magenta}\mnras}, \href{https://ui.adsabs.harvard.edu/abs/2020MNRAS.496..138B}{496, 138}

\bibitem[{{Barvainis} {et~al.}(2002){Barvainis}, {Alloin}, \& {Bremer}}]{Barvainis2002}
{Barvainis}, R., {Alloin}, D., \& {Bremer}, M. 2002, \href{http://dx.doi.org/10.1051/0004-6361:20020121}{\color{magenta}\aap}, \href{https://ui.adsabs.harvard.edu/abs/2002A&A...385..399B}{385, 399}

\bibitem[{{Barvainis} {et~al.}(1997){Barvainis}, {Maloney}, {Antonucci}, \& {Alloin}}]{Barvainis1997}
{Barvainis}, R., {Maloney}, P., {Antonucci}, R., \& {Alloin}, D. 1997, \href{http://dx.doi.org/10.1086/304382}{\color{magenta}\apj}, \href{https://ui.adsabs.harvard.edu/abs/1997ApJ...484..695B}{484, 695}

\bibitem[{{Bigiel} {et~al.}(2008){Bigiel}, {Leroy}, {Walter}, {Brinks}, {de Blok}, {Madore}, \& {Thornley}}]{Bigiel2008}
{Bigiel}, F., {Leroy}, A., {Walter}, F., {et~al.} 2008, \href{http://dx.doi.org/10.1088/0004-6256/136/6/2846}{\color{magenta}\aj}, \href{https://ui.adsabs.harvard.edu/abs/2008AJ....136.2846B}{136, 2846}

\bibitem[{{Blackburne} {et~al.}(2011){Blackburne}, {Pooley}, {Rappaport}, \& {Schechter}}]{Blackburne2011}
{Blackburne}, J.~A., {Pooley}, D., {Rappaport}, S., \& {Schechter}, P.~L. 2011, \href{http://dx.doi.org/10.1088/0004-637X/729/1/34}{\color{magenta}\apj}, \href{https://ui.adsabs.harvard.edu/abs/2011ApJ...729...34B}{729, 34}

\bibitem[{{Bothwell} {et~al.}(2013){Bothwell}, {Smail}, {Chapman}, {Genzel}, {Ivison}, {Tacconi}, {Alaghband-Zadeh}, {Bertoldi}, {Blain}, {Casey}, {Cox}, {Greve}, {Lutz}, {Neri}, {Omont}, \& {Swinbank}}]{Bothwell2013}
{Bothwell}, M.~S., {Smail}, I., {Chapman}, S.~C., {et~al.} 2013, \href{http://dx.doi.org/10.1093/mnras/sts562}{\color{magenta}\mnras}, \href{https://ui.adsabs.harvard.edu/abs/2013MNRAS.429.3047B}{429, 3047}

\bibitem[{{Bradford} {et~al.}(2009){Bradford}, {Aguirre}, {Aikin}, {Bock}, {Earle}, {Glenn}, {Inami}, {Maloney}, {Matsuhara}, {Naylor}, {Nguyen}, \& {Zmuidzinas}}]{Bradford2009}
{Bradford}, C.~M., {Aguirre}, J.~E., {Aikin}, R., {et~al.} 2009, \href{http://dx.doi.org/10.1088/0004-637X/705/1/112}{\color{magenta}\apj}, \href{https://ui.adsabs.harvard.edu/abs/2009ApJ...705..112B}{705, 112}

\bibitem[{{Bussmann} {et~al.}(2013){Bussmann}, {P{\'e}rez-Fournon}, {Amber}, {Calanog}, {Gurwell}, {Dannerbauer}, {De Bernardis}, {Fu}, {Harris}, {Krips}, {Lapi}, {Maiolino}, {Omont}, {Riechers}, {Wardlow}, {Baker}, {Birkinshaw}, {Bock}, {Bourne}, {Clements}, {Cooray}, {De Zotti}, {Dunne}, {Dye}, {Eales}, {Farrah}, {Gavazzi}, {Gonz{\'a}lez Nuevo}, {Hopwood}, {Ibar}, {Ivison}, {Laporte}, {Maddox}, {Mart{\'\i}nez-Navajas}, {Michalowski}, {Negrello}, {Oliver}, {Roseboom}, {Scott}, {Serjeant}, {Smith}, {Smith}, {Streblyanska}, {Valiante}, {van der Werf}, {Verma}, {Vieira}, {Wang}, \& {Wilner}}]{Bussmann2013}
{Bussmann}, R.~S., {P{\'e}rez-Fournon}, I., {Amber}, S., {et~al.} 2013, \href{http://dx.doi.org/10.1088/0004-637X/779/1/25}{\color{magenta}\apj}, \href{https://ui.adsabs.harvard.edu/abs/2013ApJ...779...25B}{779, 25}

\bibitem[{{Calzetti} {et~al.}(2010){Calzetti}, {Wu}, {Hong}, {Kennicutt}, {Lee}, {Dale}, {Engelbracht}, {van Zee}, {Draine}, {Hao}, {Gordon}, {Moustakas}, {Murphy}, {Regan}, {Begum}, {Block}, {Dalcanton}, {Funes}, {Gil de Paz}, {Johnson}, {Sakai}, {Skillman}, {Walter}, {Weisz}, {Williams}, \& {Wu}}]{Calzetti2010}
{Calzetti}, D., {Wu}, S.~Y., {Hong}, S., {et~al.} 2010, \href{http://dx.doi.org/10.1088/0004-637X/714/2/1256}{\color{magenta}\apj}, \href{https://ui.adsabs.harvard.edu/abs/2010ApJ...714.1256C}{714, 1256}

\bibitem[{{Carilli} \& {Walter}(2013)}]{Carilli2013}
{Carilli}, C.~L. \& {Walter}, F. 2013, \href{http://dx.doi.org/10.1146/annurev-astro-082812-140953}{\color{magenta}\araa}, \href{https://ui.adsabs.harvard.edu/abs/2013ARA&A..51..105C}{51, 105}

\bibitem[{{Carter} {et~al.}(2012){Carter}, {Lazareff}, {Maier}, {Chenu}, {Fontana}, {Bortolotti}, {Boucher}, {Navarrini}, {Blanchet}, {Greve}, {John}, {Kramer}, {Morel}, {Navarro}, {Pe{\~n}alver}, {Schuster}, \& {Thum}}]{Carter2012}
{Carter}, M., {Lazareff}, B., {Maier}, D., {et~al.} 2012, \href{http://dx.doi.org/10.1051/0004-6361/201118452}{\color{magenta}\aap}, \href{https://ui.adsabs.harvard.edu/abs/2012A&A...538A..89C}{538, A89}

\bibitem[{{Circosta} {et~al.}(2021){Circosta}, {Mainieri}, {Lamperti}, {Padovani}, {Bischetti}, {Harrison}, {Kakkad}, {Zanella}, {Vietri}, {Lanzuisi}, {Salvato}, {Brusa}, {Carniani}, {Cicone}, {Cresci}, {Feruglio}, {Husemann}, {Mannucci}, {Marconi}, {Perna}, {Piconcelli}, {Puglisi}, {Saintonge}, {Schramm}, {Vignali}, \& {Zappacosta}}]{Circosta2021}
{Circosta}, C., {Mainieri}, V., {Lamperti}, I., {et~al.} 2021, \href{http://dx.doi.org/10.1051/0004-6361/202039270}{\color{magenta}\aap}, \href{https://ui.adsabs.harvard.edu/abs/2021A&A...646A..96C}{646, A96}

\bibitem[{{Cox} {et~al.}(2002){Cox}, {Omont}, {Djorgovski}, {Bertoldi}, {Pety}, {Carilli}, {Isaak}, {Beelen}, {McMahon}, \& {Castro}}]{Cox02}
{Cox}, P., {Omont}, A., {Djorgovski}, S.~G., {et~al.} 2002, \href{http://dx.doi.org/10.1051/0004-6361:20020382}{\color{magenta}\aap}, \href{https://ui.adsabs.harvard.edu/abs/2002A&A...387..406C}{387, 406}

\bibitem[{{Crain} {et~al.}(2015){Crain}, {Schaye}, {Bower}, {Furlong}, {Schaller}, {Theuns}, {Dalla Vecchia}, {Frenk}, {McCarthy}, {Helly}, {Jenkins}, {Rosas-Guevara}, {White}, \& {Trayford}}]{Crain2015}
{Crain}, R.~A., {Schaye}, J., {Bower}, R.~G., {et~al.} 2015, \href{http://dx.doi.org/10.1093/mnras/stv725}{\color{magenta}\mnras}, \href{https://ui.adsabs.harvard.edu/abs/2015MNRAS.450.1937C}{450, 1937}

\bibitem[{{Daddi} {et~al.}(2010){Daddi}, {Bournaud}, {Walter}, {Dannerbauer}, {Carilli}, {Dickinson}, {Elbaz}, {Morrison}, {Riechers}, {Onodera}, {Salmi}, {Krips}, \& {Stern}}]{Daddi2010}
{Daddi}, E., {Bournaud}, F., {Walter}, F., {et~al.} 2010, \href{http://dx.doi.org/10.1088/0004-637X/713/1/686}{\color{magenta}\apj}, \href{https://ui.adsabs.harvard.edu/abs/2010ApJ...713..686D}{713, 686}

\bibitem[{{Daddi} {et~al.}(2009{\natexlab{a}}){Daddi}, {Dannerbauer}, {Krips}, {Walter}, {Dickinson}, {Elbaz}, \& {Morrison}}]{Daddi2009b}
{Daddi}, E., {Dannerbauer}, H., {Krips}, M., {et~al.} 2009{\natexlab{a}}, \href{http://dx.doi.org/10.1088/0004-637X/695/2/L176}{\color{magenta}\apjl}, \href{https://ui.adsabs.harvard.edu/abs/2009ApJ...695L.176D}{695, L176}

\bibitem[{{Daddi} {et~al.}(2015){Daddi}, {Dannerbauer}, {Liu}, {Aravena}, {Bournaud}, {Walter}, {Riechers}, {Magdis}, {Sargent}, {B{\'e}thermin}, {Carilli}, {Cibinel}, {Dickinson}, {Elbaz}, {Gao}, {Gobat}, {Hodge}, \& {Krips}}]{Daddi2015}
{Daddi}, E., {Dannerbauer}, H., {Liu}, D., {et~al.} 2015, \href{http://dx.doi.org/10.1051/0004-6361/201425043}{\color{magenta}\aap}, \href{https://ui.adsabs.harvard.edu/abs/2015A&A...577A..46D}{577, A46}

\bibitem[{{Daddi} {et~al.}(2009{\natexlab{b}}){Daddi}, {Dannerbauer}, {Stern}, {Dickinson}, {Morrison}, {Elbaz}, {Giavalisco}, {Mancini}, {Pope}, \& {Spinrad}}]{Daddi2009a}
{Daddi}, E., {Dannerbauer}, H., {Stern}, D., {et~al.} 2009{\natexlab{b}}, \href{http://dx.doi.org/10.1088/0004-637X/694/2/1517}{\color{magenta}\apj}, \href{https://ui.adsabs.harvard.edu/abs/2009ApJ...694.1517D}{694, 1517}

\bibitem[{{Dale} {et~al.}(2014){Dale}, {Helou}, {Magdis}, {Armus}, {D{\'\i}az-Santos}, \& {Shi}}]{Dale2014}
{Dale}, D.~A., {Helou}, G., {Magdis}, G.~E., {et~al.} 2014, \href{http://dx.doi.org/10.1088/0004-637X/784/1/83}{\color{magenta}\apj}, \href{https://ui.adsabs.harvard.edu/abs/2014ApJ...784...83D}{784, 83}

\bibitem[{{Deane} {et~al.}(2013){Deane}, {Rawlings}, {Garrett}, {Heywood}, {Jarvis}, {Kl{\"o}ckner}, {Marshall}, \& {McKean}}]{Deane2013}
{Deane}, R.~P., {Rawlings}, S., {Garrett}, M.~A., {et~al.} 2013, \href{http://dx.doi.org/10.1093/mnras/stt1241}{\color{magenta}\mnras}, \href{https://ui.adsabs.harvard.edu/abs/2013MNRAS.434.3322D}{434, 3322}

\bibitem[{{Di Mascia} {et~al.}(2023){Di Mascia}, {Carniani}, {Gallerani}, {Vito}, {Pallottini}, {Ferrara}, \& {Valentini}}]{DiMascia2023}
{Di Mascia}, F., {Carniani}, S., {Gallerani}, S., {et~al.} 2023, \href{http://dx.doi.org/10.1093/mnras/stac3306}{\color{magenta}\mnras}, \href{https://ui.adsabs.harvard.edu/abs/2023MNRAS.518.3667D}{518, 3667}

\bibitem[{{Fixsen} {et~al.}(1999){Fixsen}, {Bennett}, \& {Mather}}]{Fixsen1999}
{Fixsen}, D.~J., {Bennett}, C.~L., \& {Mather}, J.~C. 1999, \href{http://dx.doi.org/10.1086/307962}{\color{magenta}\apj}, \href{https://ui.adsabs.harvard.edu/abs/1999ApJ...526..207F}{526, 207}

\bibitem[{{Frias Castillo} {et~al.}(2024){Frias Castillo}, {Rybak}, {Hodge}, {van der Werf}, {Abbo}, {Ballieux}, {Ward}, {Harrison}, {Calistro Rivera}, {McKean}, \& {Stacey}}]{Castillo2024}
{Frias Castillo}, M., {Rybak}, M., {Hodge}, J., {et~al.} 2024, \href{http://dx.doi.org/10.1051/0004-6361/202347596}{\color{magenta}\aap}, \href{https://ui.adsabs.harvard.edu/abs/2024A&A...683A.211F}{683, A211}

\bibitem[{{Glikman} {et~al.}(2023){Glikman}, {Rusu}, {Chen}, {Chan}, {Spingola}, {Stacey}, {McKean}, {Berghea}, {Djorgovski}, {Graham}, {Stern}, {Urrutia}, {Lacy}, {Secrest}, \& {O'Meara}}]{Glikman2023}
{Glikman}, E., {Rusu}, C.~E., {Chen}, G. C.~F., {et~al.} 2023, \href{http://dx.doi.org/10.3847/1538-4357/aca093}{\color{magenta}\apj}, \href{https://ui.adsabs.harvard.edu/abs/2023ApJ...943...25G}{943, 25}

\bibitem[{{Greve} {et~al.}(2005){Greve}, {Bertoldi}, {Smail}, {Neri}, {Chapman}, {Blain}, {Ivison}, {Genzel}, {Omont}, {Cox}, {Tacconi}, \& {Kneib}}]{Greve2005}
{Greve}, T.~R., {Bertoldi}, F., {Smail}, I., {et~al.} 2005, \href{http://dx.doi.org/10.1111/j.1365-2966.2005.08979.x}{\color{magenta}\mnras}, \href{https://ui.adsabs.harvard.edu/abs/2005MNRAS.359.1165G}{359, 1165}

\bibitem[{{Grogin} {et~al.}(2005){Grogin}, {Conselice}, {Chatzichristou}, {Alexander}, {Bauer}, {Hornschemeier}, {Jogee}, {Koekemoer}, {Laidler}, {Livio}, {Lucas}, {Paolillo}, {Ravindranath}, {Schreier}, {Simmons}, \& {Urry}}]{Grogin2005}
{Grogin}, N.~A., {Conselice}, C.~J., {Chatzichristou}, E., {et~al.} 2005, \href{http://dx.doi.org/10.1086/432256}{\color{magenta}\apjl}, \href{https://ui.adsabs.harvard.edu/abs/2005ApJ...627L..97G}{627, L97}

\bibitem[{{G{\"u}rkan} {et~al.}(2015){G{\"u}rkan}, {Hardcastle}, {Jarvis}, {Smith}, {Bourne}, {Dunne}, {Maddox}, {Ivison}, \& {Fritz}}]{Gurkan15}
{G{\"u}rkan}, G., {Hardcastle}, M.~J., {Jarvis}, M.~J., {et~al.} 2015, \href{http://dx.doi.org/10.1093/mnras/stv1502}{\color{magenta}\mnras}, \href{https://ui.adsabs.harvard.edu/abs/2015MNRAS.452.3776G}{452, 3776}

\bibitem[{{Harris} {et~al.}(2016){Harris}, {Farrah}, {Schulz}, {Hatziminaoglou}, {Viero}, {Anderson}, {B{\'e}thermin}, {Chapman}, {Clements}, {Cooray}, {Efstathiou}, {Feltre}, {Hurley}, {Ibar}, {Lacy}, {Oliver}, {Page}, {P{\'e}rez-Fournon}, {Petty}, {Pitchford}, {Rigopoulou}, {Scott}, {Symeonidis}, {Vieira}, \& {Wang}}]{Harris16}
{Harris}, K., {Farrah}, D., {Schulz}, B., {et~al.} 2016, \href{http://dx.doi.org/10.1093/mnras/stw286}{\color{magenta}\mnras}, \href{https://ui.adsabs.harvard.edu/abs/2016MNRAS.457.4179H}{457, 4179}

\bibitem[{{Harrison}(2017)}]{Harrison2017}
{Harrison}, C.~M. 2017, \href{http://dx.doi.org/10.1038/s41550-017-0165}{\color{magenta}Nature Astronomy}, \href{https://ui.adsabs.harvard.edu/abs/2017NatAs...1E.165H}{1, 0165}

\bibitem[{{Ho}(2008)}]{Ho2008}
{Ho}, L.~C. 2008, \href{http://dx.doi.org/10.1146/annurev.astro.45.051806.110546}{\color{magenta}\araa}, \href{https://ui.adsabs.harvard.edu/abs/2008ARA&A..46..475H}{46, 475}

\bibitem[{{Hopkins} {et~al.}(2008){Hopkins}, {Hernquist}, {Cox}, \& {Kere{\v{s}}}}]{Hopkins2008}
{Hopkins}, P.~F., {Hernquist}, L., {Cox}, T.~J., \& {Kere{\v{s}}}, D. 2008, \href{http://dx.doi.org/10.1086/524362}{\color{magenta}\apjs}, \href{https://ui.adsabs.harvard.edu/abs/2008ApJS..175..356H}{175, 356}

\bibitem[{{Husemann} {et~al.}(2017){Husemann}, {Davis}, {Jahnke}, {Dannerbauer}, {Urrutia}, \& {Hodge}}]{Husemann2017}
{Husemann}, B., {Davis}, T.~A., {Jahnke}, K., {et~al.} 2017, \href{http://dx.doi.org/10.1093/mnras/stx1123}{\color{magenta}\mnras}, \href{https://ui.adsabs.harvard.edu/abs/2017MNRAS.470.1570H}{470, 1570}

\bibitem[{{Ivison} {et~al.}(2002){Ivison}, {Greve}, {Smail}, {Dunlop}, {Roche}, {Scott}, {Page}, {Stevens}, {Almaini}, {Blain}, {Willott}, {Fox}, {Gilbank}, {Serjeant}, \& {Hughes}}]{Ivison2002}
{Ivison}, R.~J., {Greve}, T.~R., {Smail}, I., {et~al.} 2002, \href{http://dx.doi.org/10.1046/j.1365-8711.2002.05900.x}{\color{magenta}\mnras}, \href{https://ui.adsabs.harvard.edu/abs/2002MNRAS.337....1I}{337, 1}

\bibitem[{{Kakkad} {et~al.}(2017){Kakkad}, {Mainieri}, {Brusa}, {Padovani}, {Carniani}, {Feruglio}, {Sargent}, {Husemann}, {Bongiorno}, {Bonzini}, {Piconcelli}, {Silverman}, \& {Rujopakarn}}]{Kakkad2017}
{Kakkad}, D., {Mainieri}, V., {Brusa}, M., {et~al.} 2017, \href{http://dx.doi.org/10.1093/mnras/stx726}{\color{magenta}\mnras}, \href{https://ui.adsabs.harvard.edu/abs/2017MNRAS.468.4205K}{468, 4205}

\bibitem[{{Kennicutt}(1998)}]{Kennicutt1998}
{Kennicutt}, Jr., R.~C. 1998, \href{http://dx.doi.org/10.1146/annurev.astro.36.1.189}{\color{magenta}\araa}, \href{https://ui.adsabs.harvard.edu/abs/1998ARA&A..36..189K}{36, 189}

\bibitem[{{Kormendy} \& {Ho}(2013)}]{Kormendy&Ho2013}
{Kormendy}, J. \& {Ho}, L.~C. 2013, \href{http://dx.doi.org/10.1146/annurev-astro-082708-101811}{\color{magenta}\araa}, \href{https://ui.adsabs.harvard.edu/abs/2013ARA&A..51..511K}{51, 511}

\bibitem[{{Koss} {et~al.}(2021){Koss}, {Strittmatter}, {Lamperti}, {Shimizu}, {Trakhtenbrot}, {Saintonge}, {Treister}, {Cicone}, {Mushotzky}, {Oh}, {Ricci}, {Stern}, {Ananna}, {Bauer}, {Privon}, {B{\"a}r}, {De Breuck}, {Harrison}, {Ichikawa}, {Powell}, {Rosario}, {Sanders}, {Schawinski}, {Shao}, {Megan Urry}, \& {Veilleux}}]{Koss2021}
{Koss}, M.~J., {Strittmatter}, B., {Lamperti}, I., {et~al.} 2021, \href{http://dx.doi.org/10.3847/1538-4365/abcbfe}{\color{magenta}\apjs}, \href{https://ui.adsabs.harvard.edu/abs/2021ApJS..252...29K}{252, 29}

\bibitem[{{Lapi} {et~al.}(2018){Lapi}, {Pantoni}, {Zanisi}, {Shi}, {Mancuso}, {Massardi}, {Shankar}, {Bressan}, \& {Danese}}]{Lapi2018}
{Lapi}, A., {Pantoni}, L., {Zanisi}, L., {et~al.} 2018, \href{http://dx.doi.org/10.3847/1538-4357/aab6af}{\color{magenta}\apj}, \href{https://ui.adsabs.harvard.edu/abs/2018ApJ...857...22L}{857, 22}

\bibitem[{{Lee}(2017)}]{Lee2017}
{Lee}, C.~H. 2017, \href{http://dx.doi.org/10.1051/0004-6361/201731695}{\color{magenta}\aap}, \href{https://ui.adsabs.harvard.edu/abs/2017A&A...605L...8L}{605, L8}

\bibitem[{{Leroy} {et~al.}(2009){Leroy}, {Walter}, {Bigiel}, {Usero}, {Weiss}, {Brinks}, {de Blok}, {Kennicutt}, {Schuster}, {Kramer}, {Wiesemeyer}, \& {Roussel}}]{Leroy2009}
{Leroy}, A.~K., {Walter}, F., {Bigiel}, F., {et~al.} 2009, \href{http://dx.doi.org/10.1088/0004-6256/137/6/4670}{\color{magenta}\aj}, \href{https://ui.adsabs.harvard.edu/abs/2009AJ....137.4670L}{137, 4670}

\bibitem[{{Leroy} {et~al.}(2008){Leroy}, {Walter}, {Brinks}, {Bigiel}, {de Blok}, {Madore}, \& {Thornley}}]{Leroy2008}
{Leroy}, A.~K., {Walter}, F., {Brinks}, E., {et~al.} 2008, \href{http://dx.doi.org/10.1088/0004-6256/136/6/2782}{\color{magenta}\aj}, \href{https://ui.adsabs.harvard.edu/abs/2008AJ....136.2782L}{136, 2782}

\bibitem[{{Maccagni} {et~al.}(2014){Maccagni}, {Morganti}, {Oosterloo}, \& {Mahony}}]{Maccagni2014}
{Maccagni}, F.~M., {Morganti}, R., {Oosterloo}, T.~A., \& {Mahony}, E.~K. 2014, \href{http://dx.doi.org/10.1051/0004-6361/201424334}{\color{magenta}\aap}, \href{https://ui.adsabs.harvard.edu/abs/2014A&A...571A..67M}{571, A67}

\bibitem[{{Madau} \& {Dickinson}(2014)}]{Madau14}
{Madau}, P. \& {Dickinson}, M. 2014, \href{http://dx.doi.org/10.1146/annurev-astro-081811-125615}{\color{magenta}\araa}, \href{https://ui.adsabs.harvard.edu/abs/2014ARA&A..52..415M}{52, 415}

\bibitem[{{Marconi} \& {Hunt}(2003)}]{Marconi03}
{Marconi}, A. \& {Hunt}, L.~K. 2003, \href{http://dx.doi.org/10.1086/375804}{\color{magenta}\apjl}, \href{https://ui.adsabs.harvard.edu/abs/2003ApJ...589L..21M}{589, L21}

\bibitem[{{Matsuoka} {et~al.}(2018){Matsuoka}, {Toba}, {Shidatsu}, {Ueda}, {Iwasawa}, {Terashima}, {Imanishi}, {Nagao}, {Marconi}, \& {Wang}}]{Matsuoka18}
{Matsuoka}, K., {Toba}, Y., {Shidatsu}, M., {et~al.} 2018, \href{http://dx.doi.org/10.1051/0004-6361/201833943}{\color{magenta}\aap}, \href{https://ui.adsabs.harvard.edu/abs/2018A&A...620L...3M}{620, L3}

\bibitem[{{Mazzucchelli} {et~al.}(2017){Mazzucchelli}, {Ba{\~n}ados}, {Venemans}, {Decarli}, {Farina}, {Walter}, {Eilers}, {Rix}, {Simcoe}, {Stern}, {Fan}, {Schlafly}, {De Rosa}, {Hennawi}, {Chambers}, {Greiner}, {Burgett}, {Draper}, {Kaiser}, {Kudritzki}, {Magnier}, {Metcalfe}, {Waters}, \& {Wainscoat}}]{Mazzucchelli2017}
{Mazzucchelli}, C., {Ba{\~n}ados}, E., {Venemans}, B.~P., {et~al.} 2017, \href{http://dx.doi.org/10.3847/1538-4357/aa9185}{\color{magenta}\apj}, \href{https://ui.adsabs.harvard.edu/abs/2017ApJ...849...91M}{849, 91}

\bibitem[{{Molina} {et~al.}(2023){Molina}, {Shangguan}, {Wang}, {Ho}, {Bauer}, \& {Treister}}]{Molina2023}
{Molina}, J., {Shangguan}, J., {Wang}, R., {et~al.} 2023, \href{http://dx.doi.org/10.3847/1538-4357/acc9b4}{\color{magenta}\apj}, \href{https://ui.adsabs.harvard.edu/abs/2023ApJ...950...60M}{950, 60}

\bibitem[{{Narayanan} \& {Krumholz}(2014)}]{Narayanan2014}
{Narayanan}, D. \& {Krumholz}, M.~R. 2014, \href{http://dx.doi.org/10.1093/mnras/stu834}{\color{magenta}\mnras}, \href{https://ui.adsabs.harvard.edu/abs/2014MNRAS.442.1411N}{442, 1411}

\bibitem[{{Omont} {et~al.}(2001){Omont}, {Cox}, {Bertoldi}, {McMahon}, {Carilli}, \& {Isaak}}]{Omont01}
{Omont}, A., {Cox}, P., {Bertoldi}, F., {et~al.} 2001, \href{http://dx.doi.org/10.1051/0004-6361:20010721}{\color{magenta}\aap}, \href{https://ui.adsabs.harvard.edu/abs/2001A&A...374..371O}{374, 371}

\bibitem[{{Panuzzo} {et~al.}(2010){Panuzzo}, {Rangwala}, {Rykala}, {Isaak}, {Glenn}, {Wilson}, {Auld}, {Baes}, {Barlow}, {Bendo}, {Bock}, {Boselli}, {Bradford}, {Buat}, {Castro-Rodr{\'\i}guez}, {Chanial}, {Charlot}, {Ciesla}, {Clements}, {Cooray}, {Cormier}, {Cortese}, {Davies}, {Dwek}, {Eales}, {Elbaz}, {Fulton}, {Galametz}, {Galliano}, {Gear}, {Gomez}, {Griffin}, {Hony}, {Levenson}, {Lu}, {Madden}, {O'Halloran}, {Okumura}, {Oliver}, {Page}, {Papageorgiou}, {Parkin}, {P{\'e}rez-Fournon}, {Pohlen}, {Polehampton}, {Rigby}, {Roussel}, {Sacchi}, {Sauvage}, {Schulz}, {Schirm}, {Smith}, {Spinoglio}, {Stevens}, {Srinivasan}, {Symeonidis}, {Swinyard}, {Trichas}, {Vaccari}, {Vigroux}, {Wozniak}, {Wright}, \& {Zeilinger}}]{Panuzzo2010}
{Panuzzo}, P., {Rangwala}, N., {Rykala}, A., {et~al.} 2010, \href{http://dx.doi.org/10.1051/0004-6361/201014558}{\color{magenta}\aap}, \href{https://ui.adsabs.harvard.edu/abs/2010A&A...518L..37P}{518, L37}

\bibitem[{{Papadopoulos} {et~al.}(2012){Papadopoulos}, {van der Werf}, {Xilouris}, {Isaak}, {Gao}, \& {M{\"u}hle}}]{Papadopoulos2012}
{Papadopoulos}, P.~P., {van der Werf}, P.~P., {Xilouris}, E.~M., {et~al.} 2012, \href{http://dx.doi.org/10.1111/j.1365-2966.2012.21001.x}{\color{magenta}\mnras}, \href{https://ui.adsabs.harvard.edu/abs/2012MNRAS.426.2601P}{426, 2601}

\bibitem[{{Paraficz} {et~al.}(2018){Paraficz}, {Rybak}, {McKean}, {Vegetti}, {Sluse}, {Courbin}, {Stacey}, {Suyu}, {Dessauges-Zavadsky}, {Fassnacht}, \& {Koopmans}}]{Paraficz2018}
{Paraficz}, D., {Rybak}, M., {McKean}, J.~P., {et~al.} 2018, \href{http://dx.doi.org/10.1051/0004-6361/201731250}{\color{magenta}\aap}, \href{https://ui.adsabs.harvard.edu/abs/2018A&A...613A..34P}{613, A34}

\bibitem[{{Pitchford} {et~al.}(2016){Pitchford}, {Hatziminaoglou}, {Feltre}, {Farrah}, {Clarke}, {Harris}, {Hurley}, {Oliver}, {Page}, \& {Wang}}]{Pitchford16}
{Pitchford}, L.~K., {Hatziminaoglou}, E., {Feltre}, A., {et~al.} 2016, \href{http://dx.doi.org/10.1093/mnras/stw1840}{\color{magenta}\mnras}, \href{https://ui.adsabs.harvard.edu/abs/2016MNRAS.462.4067P}{462, 4067}

\bibitem[{{Planck Collaboration} {et~al.}(2020){Planck Collaboration}, {Aghanim}, {Akrami}, {Ashdown}, {Aumont}, {Baccigalupi}, {Ballardini}, {Banday}, {Barreiro}, {Bartolo}, {Basak}, {Battye}, {Benabed}, {Bernard}, {Bersanelli}, {Bielewicz}, {Bock}, {Bond}, {Borrill}, {Bouchet}, {Boulanger}, {Bucher}, {Burigana}, {Butler}, {Calabrese}, {Cardoso}, {Carron}, {Challinor}, {Chiang}, {Chluba}, {Colombo}, {Combet}, {Contreras}, {Crill}, {Cuttaia}, {de Bernardis}, {de Zotti}, {Delabrouille}, {Delouis}, {Di Valentino}, {Diego}, {Dor{\'e}}, {Douspis}, {Ducout}, {Dupac}, {Dusini}, {Efstathiou}, {Elsner}, {En{\ss}lin}, {Eriksen}, {Fantaye}, {Farhang}, {Fergusson}, {Fernandez-Cobos}, {Finelli}, {Forastieri}, {Frailis}, {Fraisse}, {Franceschi}, {Frolov}, {Galeotta}, {Galli}, {Ganga}, {G{\'e}nova-Santos}, {Gerbino}, {Ghosh}, {Gonz{\'a}lez-Nuevo}, {G{\'o}rski}, {Gratton}, {Gruppuso}, {Gudmundsson}, {Hamann}, {Handley}, {Hansen}, {Herranz}, {Hildebrandt}, {Hivon}, {Huang}, {Jaffe}, {Jones}, {Karakci}, {Keih{\"a}nen},
  {Keskitalo}, {Kiiveri}, {Kim}, {Kisner}, {Knox}, {Krachmalnicoff}, {Kunz}, {Kurki-Suonio}, {Lagache}, {Lamarre}, {Lasenby}, {Lattanzi}, {Lawrence}, {Le Jeune}, {Lemos}, {Lesgourgues}, {Levrier}, {Lewis}, {Liguori}, {Lilje}, {Lilley}, {Lindholm}, {L{\'o}pez-Caniego}, {Lubin}, {Ma}, {Mac{\'\i}as-P{\'e}rez}, {Maggio}, {Maino}, {Mandolesi}, {Mangilli}, {Marcos-Caballero}, {Maris}, {Martin}, {Martinelli}, {Mart{\'\i}nez-Gonz{\'a}lez}, {Matarrese}, {Mauri}, {McEwen}, {Meinhold}, {Melchiorri}, {Mennella}, {Migliaccio}, {Millea}, {Mitra}, {Miville-Desch{\^e}nes}, {Molinari}, {Montier}, {Morgante}, {Moss}, {Natoli}, {N{\o}rgaard-Nielsen}, {Pagano}, {Paoletti}, {Partridge}, {Patanchon}, {Peiris}, {Perrotta}, {Pettorino}, {Piacentini}, {Polastri}, {Polenta}, {Puget}, {Rachen}, {Reinecke}, {Remazeilles}, {Renzi}, {Rocha}, {Rosset}, {Roudier}, {Rubi{\~n}o-Mart{\'\i}n}, {Ruiz-Granados}, {Salvati}, {Sandri}, {Savelainen}, {Scott}, {Shellard}, {Sirignano}, {Sirri}, {Spencer}, {Sunyaev}, {Suur-Uski}, {Tauber}, {Tavagnacco},
  {Tenti}, {Toffolatti}, {Tomasi}, {Trombetti}, {Valenziano}, {Valiviita}, {Van Tent}, {Vibert}, {Vielva}, {Villa}, {Vittorio}, {Wandelt}, {Wehus}, {White}, {White}, {Zacchei}, \& {Zonca}}]{PlanckCollaboration2020}
{Planck Collaboration}, {Aghanim}, N., {Akrami}, Y., {et~al.} 2020, \href{http://dx.doi.org/10.1051/0004-6361/201833910}{\color{magenta}\aap}, \href{https://ui.adsabs.harvard.edu/abs/2020A&A...641A...6P}{641, A6}

\bibitem[{{Popovi{\'c}}(2020)}]{Popovic2020}
{Popovi{\'c}}, L.~{\v{C}}. 2020, \href{http://dx.doi.org/10.1515/astro-2020-0003}{\color{magenta}Open Astronomy}, \href{https://ui.adsabs.harvard.edu/abs/2020OAst...29....1P}{29, 1}

\bibitem[{{Riechers}(2011)}]{Riechers2011}
{Riechers}, D.~A. 2011, \href{http://dx.doi.org/10.1088/0004-637X/730/2/108}{\color{magenta}\apj}, \href{https://ui.adsabs.harvard.edu/abs/2011ApJ...730..108R}{730, 108}

\bibitem[{{Riechers} {et~al.}(2006){Riechers}, {Walter}, {Carilli}, {Knudsen}, {Lo}, {Benford}, {Staguhn}, {Hunter}, {Bertoldi}, {Henkel}, {Menten}, {Weiss}, {Yun}, \& {Scoville}}]{Riechers2006}
{Riechers}, D.~A., {Walter}, F., {Carilli}, C.~L., {et~al.} 2006, \href{http://dx.doi.org/10.1086/507014}{\color{magenta}\apj}, \href{https://ui.adsabs.harvard.edu/abs/2006ApJ...650..604R}{650, 604}

\bibitem[{{Sabater} {et~al.}(2015){Sabater}, {Best}, \& {Heckman}}]{Sabater2015}
{Sabater}, J., {Best}, P.~N., \& {Heckman}, T.~M. 2015, \href{http://dx.doi.org/10.1093/mnras/stu2429}{\color{magenta}\mnras}, \href{https://ui.adsabs.harvard.edu/abs/2015MNRAS.447..110S}{447, 110}

\bibitem[{{Salvestrini} {et~al.}(2025){Salvestrini}, {Feruglio}, {Tripodi}, {Fontanot}, {Bischetti}, {De Lucia}, {Fiore}, {Hirschmann}, {Maio}, {Piconcelli}, {Saccheo}, {Tortosa}, {Valiante}, {Xie}, \& {Zappacosta}}]{Salvestrini2025}
{Salvestrini}, F., {Feruglio}, C., {Tripodi}, R., {et~al.} 2025, \href{http://dx.doi.org/10.1051/0004-6361/202453226}{\color{magenta}\aap}, \href{https://ui.adsabs.harvard.edu/abs/2025A&A...695A..23S}{695, A23}

\bibitem[{{Salvestrini} {et~al.}(2022){Salvestrini}, {Gruppioni}, {Hatziminaoglou}, {Pozzi}, {Vignali}, {Casasola}, {Paladino}, {Aalto}, {Andreani}, {Marchesi}, \& {Stanke}}]{Salvestrini2022}
{Salvestrini}, F., {Gruppioni}, C., {Hatziminaoglou}, E., {et~al.} 2022, \href{http://dx.doi.org/10.1051/0004-6361/202142760}{\color{magenta}\aap}, \href{https://ui.adsabs.harvard.edu/abs/2022A&A...663A..28S}{663, A28}

\bibitem[{{Sanders} \& {Mirabel}(1985)}]{Sanders1985}
{Sanders}, D.~B. \& {Mirabel}, I.~F. 1985, \href{http://dx.doi.org/10.1086/184561}{\color{magenta}\apjl}, \href{https://ui.adsabs.harvard.edu/abs/1985ApJ...298L..31S}{298, L31}

\bibitem[{{Sanders} {et~al.}(1988){Sanders}, {Soifer}, {Elias}, {Madore}, {Matthews}, {Neugebauer}, \& {Scoville}}]{Sanders1988}
{Sanders}, D.~B., {Soifer}, B.~T., {Elias}, J.~H., {et~al.} 1988, \href{http://dx.doi.org/10.1086/165983}{\color{magenta}\apj}, \href{https://ui.adsabs.harvard.edu/abs/1988ApJ...325...74S}{325, 74}

\bibitem[{{Sargent} {et~al.}(2014){Sargent}, {Daddi}, {B{\'e}thermin}, {Aussel}, {Magdis}, {Hwang}, {Juneau}, {Elbaz}, \& {da Cunha}}]{Sargent2014}
{Sargent}, M.~T., {Daddi}, E., {B{\'e}thermin}, M., {et~al.} 2014, \href{http://dx.doi.org/10.1088/0004-637X/793/1/19}{\color{magenta}\apj}, \href{https://ui.adsabs.harvard.edu/abs/2014ApJ...793...19S}{793, 19}

\bibitem[{{Shangguan} {et~al.}(2020){Shangguan}, {Ho}, {Bauer}, {Wang}, \& {Treister}}]{Shangguan2020}
{Shangguan}, J., {Ho}, L.~C., {Bauer}, F.~E., {Wang}, R., \& {Treister}, E. 2020, \href{http://dx.doi.org/10.3847/1538-4365/ab5db2}{\color{magenta}\apjs}, \href{https://ui.adsabs.harvard.edu/abs/2020ApJS..247...15S}{247, 15}

\bibitem[{{Shangguan} {et~al.}(2018){Shangguan}, {Ho}, \& {Xie}}]{Shangguan18}
{Shangguan}, J., {Ho}, L.~C., \& {Xie}, Y. 2018, \href{http://dx.doi.org/10.3847/1538-4357/aaa9be}{\color{magenta}\apj}, \href{https://ui.adsabs.harvard.edu/abs/2018ApJ...854..158S}{854, 158}

\bibitem[{{Shankar} {et~al.}(2009){Shankar}, {Weinberg}, \& {Miralda-Escud{\'e}}}]{Shankar09}
{Shankar}, F., {Weinberg}, D.~H., \& {Miralda-Escud{\'e}}, J. 2009, \href{http://dx.doi.org/10.1088/0004-637X/690/1/20}{\color{magenta}\apj}, \href{https://ui.adsabs.harvard.edu/abs/2009ApJ...690...20S}{690, 20}

\bibitem[{{Sharma} {et~al.}(2024){Sharma}, {Choi}, {Somerville}, {Snyder}, {Jhee}, {Kocevski}, {Hirschmann}, {Moster}, {Naab}, {Narayanan}, {Ostriker}, \& {Rosario}}]{Sharma2024}
{Sharma}, R.~S., {Choi}, E., {Somerville}, R.~S., {et~al.} 2024, \href{http://dx.doi.org/10.1093/mnras/stad3836}{\color{magenta}\mnras}, \href{https://ui.adsabs.harvard.edu/abs/2024MNRAS.527.9461S}{527, 9461}

\bibitem[{{Sharon} {et~al.}(2019){Sharon}, {Tagore}, {Baker}, {Rivera}, {Keeton}, {Lutz}, {Genzel}, {Wilner}, {Hicks}, {Allam}, \& {Tucker}}]{Sharon2019}
{Sharon}, C.~E., {Tagore}, A.~S., {Baker}, A.~J., {et~al.} 2019, \href{http://dx.doi.org/10.3847/1538-4357/ab22b9}{\color{magenta}\apj}, \href{https://ui.adsabs.harvard.edu/abs/2019ApJ...879...52S}{879, 52}

\bibitem[{{Shen} {et~al.}(2008){Shen}, {Greene}, {Strauss}, {Richards}, \& {Schneider}}]{Shen2008}
{Shen}, Y., {Greene}, J.~E., {Strauss}, M.~A., {Richards}, G.~T., \& {Schneider}, D.~P. 2008, \href{http://dx.doi.org/10.1086/587475}{\color{magenta}\apj}, \href{https://ui.adsabs.harvard.edu/abs/2008ApJ...680..169S}{680, 169}

\bibitem[{{Shi} {et~al.}(2007){Shi}, {Ogle}, {Rieke}, {Antonucci}, {Hines}, {Smith}, {Low}, {Bouwman}, \& {Willmer}}]{Shi2007}
{Shi}, Y., {Ogle}, P., {Rieke}, G.~H., {et~al.} 2007, \href{http://dx.doi.org/10.1086/521594}{\color{magenta}\apj}, \href{https://ui.adsabs.harvard.edu/abs/2007ApJ...669..841S}{669, 841}

\bibitem[{{Shi} {et~al.}(2014{\natexlab{a}}){Shi}, {Rieke}, {Ogle}, {Su}, \& {Balog}}]{Shi14}
{Shi}, Y., {Rieke}, G.~H., {Ogle}, P.~M., {Su}, K.~Y.~L., \& {Balog}, Z. 2014{\natexlab{a}}, \href{http://dx.doi.org/10.1088/0067-0049/214/2/23}{\color{magenta}\apjs}, \href{https://ui.adsabs.harvard.edu/abs/2014ApJS..214...23S}{214, 23}

\bibitem[{{Shi} {et~al.}(2014{\natexlab{b}}){Shi}, {Rieke}, {Ogle}, {Su}, \& {Balog}}]{Shi2014}
{Shi}, Y., {Rieke}, G.~H., {Ogle}, P.~M., {Su}, K.~Y.~L., \& {Balog}, Z. 2014{\natexlab{b}}, \href{http://dx.doi.org/10.1088/0067-0049/214/2/23}{\color{magenta}\apjs}, \href{https://ui.adsabs.harvard.edu/abs/2014ApJS..214...23S}{214, 23}

\bibitem[{{Sluse} {et~al.}(2012){Sluse}, {Hutsem{\'e}kers}, {Courbin}, {Meylan}, \& {Wambsganss}}]{Sluse2012}
{Sluse}, D., {Hutsem{\'e}kers}, D., {Courbin}, F., {Meylan}, G., \& {Wambsganss}, J. 2012, \href{http://dx.doi.org/10.1051/0004-6361/201219125}{\color{magenta}\aap}, \href{https://ui.adsabs.harvard.edu/abs/2012A&A...544A..62S}{544, A62}

\bibitem[{{Sluse} {et~al.}(2003){Sluse}, {Surdej}, {Claeskens}, {Hutsem{\'e}kers}, {Jean}, {Courbin}, {Nakos}, {Billeres}, \& {Khmil}}]{Sluse2003}
{Sluse}, D., {Surdej}, J., {Claeskens}, J.~F., {et~al.} 2003, \href{http://dx.doi.org/10.1051/0004-6361:20030904}{\color{magenta}\aap}, \href{https://ui.adsabs.harvard.edu/abs/2003A&A...406L..43S}{406, L43}

\bibitem[{{Solomon} {et~al.}(2003){Solomon}, {Vanden Bout}, {Carilli}, \& {Guelin}}]{Solomon2003}
{Solomon}, P., {Vanden Bout}, P., {Carilli}, C., \& {Guelin}, M. 2003, \href{http://dx.doi.org/10.1038/nature02149}{\color{magenta}\nat}, \href{https://ui.adsabs.harvard.edu/abs/2003Natur.426..636S}{426, 636}

\bibitem[{{Solomon} {et~al.}(1997){Solomon}, {Downes}, {Radford}, \& {Barrett}}]{Solomon1997}
{Solomon}, P.~M., {Downes}, D., {Radford}, S.~J.~E., \& {Barrett}, J.~W. 1997, \href{http://dx.doi.org/10.1086/303765}{\color{magenta}\apj}, \href{https://ui.adsabs.harvard.edu/abs/1997ApJ...478..144S}{478, 144}

\bibitem[{{Solomon} \& {Vanden Bout}(2005)}]{Solomon2005}
{Solomon}, P.~M. \& {Vanden Bout}, P.~A. 2005, \href{http://dx.doi.org/10.1146/annurev.astro.43.051804.102221}{\color{magenta}\araa}, \href{https://ui.adsabs.harvard.edu/abs/2005ARA&A..43..677S}{43, 677}

\bibitem[{{Somerville} {et~al.}(2008){Somerville}, {Hopkins}, {Cox}, {Robertson}, \& {Hernquist}}]{Somerville2008}
{Somerville}, R.~S., {Hopkins}, P.~F., {Cox}, T.~J., {Robertson}, B.~E., \& {Hernquist}, L. 2008, \href{http://dx.doi.org/10.1111/j.1365-2966.2008.13805.x}{\color{magenta}\mnras}, \href{https://ui.adsabs.harvard.edu/abs/2008MNRAS.391..481S}{391, 481}

\bibitem[{{Stacey} {et~al.}(2022){Stacey}, {Costa}, {McKean}, {Sharon}, {Calistro Rivera}, {Glikman}, \& {van der Werf}}]{Stacey2022}
{Stacey}, H.~R., {Costa}, T., {McKean}, J.~P., {et~al.} 2022, \href{http://dx.doi.org/10.1093/mnras/stac2765}{\color{magenta}\mnras}, \href{https://ui.adsabs.harvard.edu/abs/2022MNRAS.517.3377S}{517, 3377}

\bibitem[{{Stacey} {et~al.}(2020){Stacey}, {Lafontaine}, \& {McKean}}]{Stacey2020}
{Stacey}, H.~R., {Lafontaine}, A., \& {McKean}, J.~P. 2020, \href{http://dx.doi.org/10.1093/mnras/staa494}{\color{magenta}\mnras}, \href{https://ui.adsabs.harvard.edu/abs/2020MNRAS.493.5290S}{493, 5290}

\bibitem[{{Stacey} {et~al.}(2021){Stacey}, {McKean}, {Powell}, {Vegetti}, {Rizzo}, {Spingola}, {Auger}, {Ivison}, \& {van der Werf}}]{Stacey2021}
{Stacey}, H.~R., {McKean}, J.~P., {Powell}, D.~M., {et~al.} 2021, \href{http://dx.doi.org/10.1093/mnras/staa3433}{\color{magenta}\mnras}, \href{https://ui.adsabs.harvard.edu/abs/2021MNRAS.500.3667S}{500, 3667}

\bibitem[{{Stacey} {et~al.}(2018){Stacey}, {McKean}, {Robertson}, {Ivison}, {Isaak}, {Schleicher}, {van der Werf}, {Baan}, {Berciano Alba}, {Garrett}, \& {Loenen}}]{Stacey2018}
{Stacey}, H.~R., {McKean}, J.~P., {Robertson}, N.~C., {et~al.} 2018, \href{http://dx.doi.org/10.1093/mnras/sty458}{\color{magenta}\mnras}, \href{https://ui.adsabs.harvard.edu/abs/2018MNRAS.476.5075S}{476, 5075}

\bibitem[{{Stanley} {et~al.}(2015){Stanley}, {Harrison}, {Alexander}, {Swinbank}, {Aird}, {Del Moro}, {Hickox}, \& {Mullaney}}]{Stanley15}
{Stanley}, F., {Harrison}, C.~M., {Alexander}, D.~M., {et~al.} 2015, \href{http://dx.doi.org/10.1093/mnras/stv1678}{\color{magenta}\mnras}, \href{https://ui.adsabs.harvard.edu/abs/2015MNRAS.453..591S}{453, 591}

\bibitem[{{Storchi-Bergmann} \& {Schnorr-M{\"u}ller}(2019)}]{Bergmann19}
{Storchi-Bergmann}, T. \& {Schnorr-M{\"u}ller}, A. 2019, \href{http://dx.doi.org/10.1038/s41550-018-0611-0}{\color{magenta}Nature Astronomy}, \href{https://ui.adsabs.harvard.edu/abs/2019NatAs...3...48S}{3, 48}

\bibitem[{{Tacconi} {et~al.}(2018){Tacconi}, {Genzel}, {Saintonge}, {Combes}, {Garc{\'\i}a-Burillo}, {Neri}, {Bolatto}, {Contini}, {F{\"o}rster Schreiber}, {Lilly}, {Lutz}, {Wuyts}, {Accurso}, {Boissier}, {Boone}, {Bouch{\'e}}, {Bournaud}, {Burkert}, {Carollo}, {Cooper}, {Cox}, {Feruglio}, {Freundlich}, {Herrera-Camus}, {Juneau}, {Lippa}, {Naab}, {Renzini}, {Salome}, {Sternberg}, {Tadaki}, {{\"U}bler}, {Walter}, {Weiner}, \& {Weiss}}]{Tacconi2018}
{Tacconi}, L.~J., {Genzel}, R., {Saintonge}, A., {et~al.} 2018, \href{http://dx.doi.org/10.3847/1538-4357/aaa4b4}{\color{magenta}\apj}, \href{https://ui.adsabs.harvard.edu/abs/2018ApJ...853..179T}{853, 179}

\bibitem[{{Tuan-Anh} {et~al.}(2017){Tuan-Anh}, {Hoai}, {Nhung}, {Diep}, {Phuong}, {Thao}, \& {Darriulat}}]{Tuan2017}
{Tuan-Anh}, P., {Hoai}, D.~T., {Nhung}, P.~T., {et~al.} 2017, \href{http://dx.doi.org/10.1093/mnras/stx212}{\color{magenta}\mnras}, \href{https://ui.adsabs.harvard.edu/abs/2017MNRAS.467.3513T}{467, 3513}

\bibitem[{{Tung} \& {Chen}(2025)}]{Tung2025}
{Tung}, P.-C. \& {Chen}, K.-J. 2025, \href{http://dx.doi.org/10.3847/1538-4357/ade1d4}{\color{magenta}\apj}, \href{https://ui.adsabs.harvard.edu/abs/2025ApJ...988..127T}{988, 127}

\bibitem[{{Villforth}(2023)}]{Villforth2023}
{Villforth}, C. 2023, \href{http://dx.doi.org/10.21105/astro.2309.03276}{\color{magenta}The Open Journal of Astrophysics}, \href{https://ui.adsabs.harvard.edu/abs/2023OJAp....6E..34V}{6, 34}

\bibitem[{{Walton} {et~al.}(2022){Walton}, {Reynolds}, {Stern}, {Brightman}, \& {Lemon}}]{Walton2022}
{Walton}, D.~J., {Reynolds}, M.~T., {Stern}, D., {Brightman}, M., \& {Lemon}, C. 2022, \href{http://dx.doi.org/10.1093/mnras/stac2554}{\color{magenta}\mnras}, \href{https://ui.adsabs.harvard.edu/abs/2022MNRAS.516.5997W}{516, 5997}

\bibitem[{{Wei{\ss}} {et~al.}(2005){Wei{\ss}}, {Walter}, \& {Scoville}}]{Weib2005}
{Wei{\ss}}, A., {Walter}, F., \& {Scoville}, N.~Z. 2005, \href{http://dx.doi.org/10.1051/0004-6361:20052667}{\color{magenta}\aap}, \href{https://ui.adsabs.harvard.edu/abs/2005A&A...438..533W}{438, 533}

\bibitem[{{Wen} \& {Kemball}(2022)}]{Wen2022arXiv}
{Wen}, D. \& {Kemball}, A.~J. 2022, \href{https://ui.adsabs.harvard.edu/abs/2022arXiv221016444W}{\href{http://dx.doi.org/10.48550/arXiv.2210.16444}{\color{magenta}arXiv e-prints, submitted to ApJ}, arXiv:2210.16444}

\bibitem[{{Wilson} {et~al.}(2009){Wilson}, {Warren}, {Israel}, {Serjeant}, {Bendo}, {Brinks}, {Clements}, {Courteau}, {Irwin}, {Knapen}, {Leech}, {Matthews}, {M{\"u}hle}, {Mortier}, {Petitpas}, {Sinukoff}, {Spekkens}, {Tan}, {Tilanus}, {Usero}, {van der Werf}, {Wiegert}, \& {Zhu}}]{Wilson2009}
{Wilson}, C.~D., {Warren}, B.~E., {Israel}, F.~P., {et~al.} 2009, \href{http://dx.doi.org/10.1088/0004-637X/693/2/1736}{\color{magenta}\apj}, \href{https://ui.adsabs.harvard.edu/abs/2009ApJ...693.1736W}{693, 1736}

\bibitem[{{Zhang} {et~al.}(2023){Zhang}, {Zhang}, {Nightingale}, {Zou}, {Cao}, {Tsai}, {Yang}, {Shi}, {Wang}, {Xu}, {Lin}, {Zhou}, \& {Li}}]{Zhang2023}
{Zhang}, L., {Zhang}, Z.-Y., {Nightingale}, J.~W., {et~al.} 2023, \href{http://dx.doi.org/10.1093/mnras/stad2069}{\color{magenta}\mnras}, \href{https://ui.adsabs.harvard.edu/abs/2023MNRAS.524.3671Z}{524, 3671}

\bibitem[{{Zhang} {et~al.}(2016){Zhang}, {Shi}, {Rieke}, {Xia}, {Wang}, {Sun}, \& {Wan}}]{Zhang16}
{Zhang}, Z., {Shi}, Y., {Rieke}, G.~H., {et~al.} 2016, \href{http://dx.doi.org/10.3847/2041-8205/819/2/L27}{\color{magenta}\apjl}, \href{https://ui.adsabs.harvard.edu/abs/2016ApJ...819L..27Z}{819, L27}

\end{thebibliography}
\end{document}